\documentclass[journal]{IEEEtran}

\usepackage{multirow}
\usepackage[table,xcdraw]{xcolor}
\usepackage{graphicx}
\usepackage{amsmath}
\usepackage{amsfonts}
\usepackage{algorithm}
\usepackage{amsthm,amssymb}
\usepackage{mathrsfs}
\usepackage{algorithmic}
\usepackage{makecell}
\usepackage{easyReview}
\usepackage{booktabs}
\setcellgapes{3pt}
\usepackage{verbatim}
\setcellgapes{3pt}

\usepackage[justification=centering]{caption}
\usepackage{color}
\usepackage[subfigure]{tocloft}
\usepackage{subfigure}
\usepackage{tabularx}
\usepackage{amsmath, bm}
\usepackage{hyperref}
\hypersetup{hidelinks,
	colorlinks=true,
	allcolors=blue,
	pdfstartview=Fit,
	breaklinks=true}

\usepackage{cite}

\ifCLASSINFOpdf
\else
\fi
\begin{document}
	\title{Large Language Model-Driven Distributed Integrated Multimodal Sensing and Semantic Communications}
	
	\author{Yubo Peng, \textit{Student Member, IEEE}, Luping Xiang, \textit{Senior Member, IEEE}, Bingxin Zhang, \textit{Member, IEEE}, and Kun Yang, \textit{Fellow, IEEE}
		\thanks{
			{Yubo Peng (ybpeng@smail.nju.edu.cn), Luping Xiang (luping.xiang@nju.edu.cn), Bingxin Zhang (bxzhang@nju.edu.cn), and Kun Yang (kunyang@essex.ac.uk) are with the State Key Laboratory of Novel Software Technology, Nanjing University, Nanjing, China, and the School of Intelligent Software and Engineering, Nanjing University (Suzhou Campus), Suzhou, China.

			}
		}
	}

\markboth{Submitted for Review}%
{Shell \MakeLowercase{\textit{et al.}}: Bare Demo of IEEEtran.cls for IEEE Journals}

\maketitle 

\begin{abstract}
Traditional single-modal sensing systems-based solely on either radio frequency (RF) or visual data-struggle to cope with the demands of complex and dynamic environments. Furthermore, single-device systems are constrained by limited perspectives and insufficient spatial coverage, which impairs their effectiveness in urban or non-line-of-sight scenarios. To overcome these challenges, we propose a novel large language model (LLM)-driven distributed integrated multimodal sensing and semantic communication (LLM-DiSAC) framework. Specifically, our system consists of multiple collaborative sensing devices equipped with RF and camera modules, working together with an aggregation center to enhance sensing accuracy. 
First, on sensing devices, LLM-DiSAC develops an RF-vision fusion network (RVFN), which employs specialized feature extractors for RF and visual data, followed by a cross-attention module for effective multimodal integration. 
Second, a LLM-based semantic transmission network (LSTN) is proposed to enhance communication efficiency, where the LLM-based decoder leverages known channel parameters, such as transceiver distance and signal-to-noise ratio (SNR), to mitigate semantic distortion. 
Third, at the aggregation center, a transformer-based aggregation model (TRAM) with an adaptive aggregation attention mechanism is developed to fuse distributed features and enhance sensing accuracy. To preserve data privacy, a two-stage distributed learning strategy is introduced, allowing local model training at the device level and centralized aggregation model training using intermediate features. Finally, evaluations on a synthetic multi-view RF-visual dataset generated by the Genesis simulation engine show that LLM-DiSAC achieves a 191.0\% relative improvement in classification accuracy, reduces RMSE by an average of 31.5\%, and lowers NMSE by 55.6\%, compared to the unimodal single-device baseline. Moreover, the data transmission cost is reduced by 92.6\% relative to conventional approaches.
\end{abstract}


%
\IEEEpeerreviewmaketitle

\section{Introduction}
\subsection{Sensing}
Sensing technologies have gained widespread adoption due to their significant advantages in various applications, such as autonomous driving, virtual reality, and beyond \cite{10599121,10878446}. However, single-modal sensing systems relying solely on radio frequency (RF) or visual data exhibit notable limitations in complex and dynamic environments. For instance, RF-based sensing, while capable of operating around the clock and providing high-ranging accuracy, fails to infer the identification or class of monitored targets, which is critical for adaptive communication strategies. Conversely, visual sensing, despite its ability to extract rich image features, suffers from severe performance degradation under adverse weather conditions, such as rain, snow, fog, and haze. Additionally, visual systems struggle to accurately measure the speed of high-velocity targets, such as vehicles on a highway, limiting their effectiveness in real-time applications, like autonomous driving \cite{peng2025simac}.

By combining complementary information from diverse sources, such as RF and visual data, multimodal sensing systems provide significant advantages over single-modal approaches. For example, in high-reliability Internet of Vehicles (IoV) scenarios, visual sensing can identify accident-prone vehicles ahead through feature extraction, while RF sensing can predict signal occlusion periods \cite{10908658,deng2025fusegrasp}. This integration enables dynamic adjustment of beamforming strategies, supporting real-time decision-making for Level 4 (L4) autonomous driving. 
However, single-device sensing systems are inherently constrained by their fixed perspectives and limited coverage. For example, a camera mounted on a vehicle or drone has a restricted field of view, making it difficult to detect targets that are occluded by obstacles or located at a distance. In urban environments, buildings or other structures can block the line of sight, leading to incomplete or inaccurate visual sensing of pedestrians or other vehicles \cite{10411947}. Similarly, RF-based sensing devices may struggle to detect targets in non-line-of-sight (NLOS) scenarios, such as those hidden behind walls or in dense foliage. These limitations hinder the reliability of single-device sensing systems in large-scale or complex scenarios \cite{6963485}.  

As a latent solution, distributed collaborative sensing can leverage the collective capabilities of distributed devices, offering broader sensing coverage, enhanced sensing performance, and improved sensing stability \cite{10908649}. However, traditional collaborative sensing methods often incur significant communication overhead due to frequent information exchange and rising concerns about data privacy \cite{9810792}. Furthermore, traditional aggregated strategies based on discriminative artificial intelligence (AI) models exhibit limitations in effectively integrating multimodal features, primarily due to their restricted capacity to capture complex relationships and high-dimensional data \cite{10558819,10638533}.

In summary, the current challenges in sensing technologies can be outlined as follows:  
\begin{enumerate}
    \item \textit{Limited Modal:}  
    Existing sensing systems typically rely on a single modality—either RF or visual—which inherently restricts their ability to perceive complex and dynamic environments. RF-based systems, while effective in estimating physical parameters such as range and velocity, are insufficient for capturing high-level semantic cues like object categories, interactions, or intentions. In contrast, visual sensors provide rich contextual information but are highly vulnerable to environmental factors such as poor lighting, fog, or occlusions. 
    
    \item \textit{Limited Range:}  
    The deployment of individual sensing devices imposes fundamental limitations on spatial coverage and viewpoint diversity. Sensors mounted on a single platform, such as a vehicle or robot, are constrained by their line-of-sight and fixed field of view, making them prone to blind spots and occlusion failures. For instance, a monocular camera may fail to detect targets hidden behind obstacles or located outside its narrow observation cone, while a single RF antenna may perform poorly in NLOS or cluttered environments. 

    \item \textit{Inefficient Feature Aggregation and Low Communication Efficiency:}
    Traditional aggregated methods using discriminative AI models are inadequate for integrating multi-device data, due to the inherent shortcomings of learning capacity. Additionally, frequent data exchange in distributed collaborative systems incurs high communication overhead, raising concerns about energy consumption, latency, and data privacy.  
\end{enumerate}

These challenges highlight the need for advanced solutions that balance efficiency, privacy, and performance in collaborative multimodal sensing systems.

\subsection{Semantic Communication and Large Language Model}
Semantic communication (SC) represents a transformative paradigm in information transmission, focusing on the exchange of meaningful and task-relevant information rather than raw data \cite{9679803,9450827}. Unlike traditional communication systems that prioritize bit-level accuracy, SC extracts and transmits only the essential semantic features required for specific tasks, such as image recognition or decision-making \cite{10670195}. This approach significantly reduces data redundancy, enhances transmission efficiency, and improves robustness in noisy or bandwidth-constrained environments \cite{10159023}. 

Large language models (LLMs) represent a prominent subcategory of Large AI Models (LAMs), characterized by their exceptional capacity to understand, generate, and manipulate human language across a wide range of tasks. Built on architectures such as Transformer, LLMs like ChatGPT \cite{yang2023chatgpt} and LLaMA \cite{touvron2023llama} typically contain billions of parameters and are pre-trained on massive text corpora. This enables them to capture rich semantic representations and contextual dependencies, allowing for remarkable generalization across tasks such as text summarization, question answering, code generation, and dialogue systems. In the context of future communication systems, LLMs offer powerful capabilities for semantic understanding and language-driven control, enabling more intelligent, adaptive, and human-centric interfaces. Moreover, by leveraging their large-scale pretraining, LLMs reduce the need for extensive task-specific data, facilitating rapid adaptation to dynamic environments with minimal fine-tuning \cite{10648594}.

\subsection{Our Contributions}
Based on SC and LLM, we propose a novel LLM-driven distributed integrated multimodal sensing and SC (LLM-DiSAC) framework to address the identified challenges. Specifically, we consider a distributed collective sensing scenario, which has numerous sensing devices to perform the sensing task. These devices include multiple collaborative devices and an aggregation center. The collaborative devices are responsible for collecting and preprocessing environmental information, which is then transmitted to the aggregation center to enhance its sensing accuracy. The main contributions of this work are summarized as follows:
\begin{enumerate}
    \item We introduce an RF-vision bimodal fusion network (RVFN) on each sensing device, which incorporates a hybrid visual feature extraction architecture. This design combines elements of the vision transformer (ViT) and convolutional neural networks (CNNs), thus achieving performance comparable to ViT while retaining the efficient inference capabilities of CNNs. Additionally, a specialized complex feature extractor is developed for RF signals, enabling direct processing of complex-valued inputs. To integrate these unimodal features, we introduce a cross-attention mechanism, which fuses RF and visual representations to generate comprehensive multimodal features. This approach fully leverages the complementary strengths of RF and visual modalities.

    \item We design an LLM-based semantic transmission network (LSTN) to facilitate information interaction between the collaborative devices and the aggregation center. On the collaborative devices, a semantic encoder encodes the extracted bimodal features for wireless transmission. On the aggregation center, a semantic decoder decodes the received semantics. To improve the adaptability of the channel, the semantic decoder, built on an LLM, incorporates known physical information, such as the transceiver distance and signal-to-noise ratio (SNR). This approach not only reduces communication overhead by transmitting semantic information, but also improves decoding accuracy by leveraging known physical information through the LLM.

    \item We propose a transformer-based generative aggregation model (TRAM) deployed on the aggregation center, which devises a cross-aggregation attention mechanism to adaptively integrate feature information from varying numbers of collaborative sensing devices and extract the most valuable latent features, thereby enhancing the sensing accuracy of the aggregation center. This method can effectively compensate for the insufficient perception of a single device due to its viewing angle.

    \item Considering data privacy concerns, we propose a novel two-stage distributed learning strategy. In the first stage, each sensing device independently trains the RVFN and LSTN models locally. In the second stage, only the intermediate feature is transmitted to the aggregation center, where the TRAM is trained. 
    To evaluate the efficacy of our proposed methods, we construct a synthetic multi-view RF-visual sensing dataset using Genesis \cite{Genesis}, an AI-powered physical engine. Experimental results demonstrate that LLM-DiSAC achieves a 191.0\% relative improvement in classification accuracy, reduces RMSE by an average of 31.5\%, and lowers NMSE by 55.6\%, compared to the unimodal single-device baseline. Moreover, the data transmission cost is reduced by 92.6\% relative to conventional approaches. This validates its effectiveness in addressing the challenges of multimodal sensing and SC.
\end{enumerate}

\subsection{Organization}
The rest of the paper has the following structure: Section II introduces the related works, and Section III provides a detailed description of the system model. 
Section IV presents the proposed LLM-DiSAC framework, including the implementation of the RVFN, TRAM, and LSTN modules. 
Section V employs experimental simulations to evaluate the performance of the proposed methods and discusses the advantages of multimodal fusion, LLM-based SC, and multi-device collaboration. 
Lastly, Section VI concludes this paper. Table \ref{tab:notations} summarizes the description of the variables in this paper.

\begin{table*}[htbp] 
\centering
\caption{List of Notations}
\label{tab:notations}
\renewcommand{\arraystretch}{1.2}
\begin{tabular}{clcl}
\toprule
\textbf{Symbol} & \textbf{Description} & \textbf{Symbol} & \textbf{Description} \\
\midrule
$K$ & Number of IoT devices & $N$ & Number of sensing targets \\
$B_k$ & Bandwidth for $k$-th IoTD & $P_k$ & Transmission power of $k$-th IoTD \\
$V_k$ & Achievable transmission rate & $T_k^{\mathrm{exe}}$ & Total execution time \\
$t_k^{\mathrm{ft}}$ & Feature extraction time & $t_k^{\mathrm{se}}$ & Semantic encoding time \\
$t_k^{\mathrm{sd}}$ & Semantic decoding time & $t_k^{\mathrm{fa}}$ & Feature aggregation time \\
$\mathcal{D}_k$ & Dataset of $k$-th IoTD & $Z(\cdot)$ & Bit-length function \\
$f_c$ & Central frequency of radar & $c$ & Speed of light ($3\times10^8$ m/s) \\
$\lambda$ & Wavelength of radar signal & $\Delta t$ & Sampling interval \\
$F_s$ & Sampling frequency & $T_r$ & Pulse repetition interval \\
$\phi_{k,n}$ & Azimuth angle & $\theta_{k,n}$ & Pitch angle \\
$d_{k,n}$ & Distance & $v_{k,n}$ & Radial velocity \\
$\tau_{k,n}$ & Time delay & $\mu_{k,n}$ & Doppler frequency shift \\
$\epsilon_n$ & Radar cross-section & $\mathbf{a}(\theta,\phi)$ & Steering vector \\
$\mathbf{T}$ & Time sampling sequence & $\mathbf{N}_0$ & Additive White Gaussian Noise \\
$\mathbf{H}_k$ & Channel gain & $\bm{\epsilon}$ & Transmission noise \\
$\mathbf{R}_{k,n}$ & Received radar echo & $\mathbf{V}_k$ & Visual image \\
$\mathbf{z}_{k,n}^{\mathrm{real}}$ & Real RF features & $\mathbf{z}_{k,n}^{\mathrm{imag}}$ & Imaginary RF features \\
$\mathbf{z}_{k,n,3}$ & Concatenated RF features & $\mathbf{f}_i$ & Vision feature maps \\
$\mathbf{s}_{k,n}^{\mathrm{RF}}$ & RF sensing feature & $\mathbf{s}_{k,n}^{\mathrm{CV}}$ & Vision sensing feature \\
$\mathbf{s}_{k,n}^{\mathrm{mul}}$ & Fused multimodal feature & $\mathbf{z}_{k,n}^{\mathrm{RF}}$ & RF-to-visual attention \\
$\mathbf{z}_{k,n}^{\mathrm{CV}}$ & Visual-to-RF attention & $\mathbf{z}_{k,n}^{\mathrm{fusion}}$ & Fused attention feature \\
$\mathbf{e}_{k,n}$ & Semantic encoding & $\mathbf{c}_{k,n}$ & Complex transmission symbols \\
$\mathbf{y}_{k,n}$ & Received symbols & $\hat{\mathbf{e}}_{k,n}$ & Received semantic encoding \\
$\hat{\mathbf{s}}_{k,n}^{\mathrm{mul}}$ & Reconstructed feature & $\mathbf{S}_n^{\mathrm{agg}}$ & Aggregated feature \\
$\mathbf{o}_{\mathrm{agg},n}$ & Final sensing output & $\hat{\mathbf{I}}_{\mathrm{agg},n}$ & Predicted class \\
$\hat{d}_{\mathrm{agg},n}$ & Predicted distance & $\hat{v}_{\mathrm{agg},n}$ & Predicted velocity \\
$\hat{\phi}_{\mathrm{agg},n}$ & Predicted azimuth & $\hat{\theta}_{\mathrm{agg},n}$ & Predicted pitch \\
$F_{\mathrm{sd}}(\cdot)$ & Semantic decoder & $\mathrm{DS}_i(\cdot)$ & Downsampling operation \\
$\mathcal{B}_i$ & ConvNeXt blocks & $\mathcal{C}_{k,n}$ & Channel context \\
$\mathbf{W}_q^1,\mathbf{W}_k^1,\mathbf{W}_v^1$ & RF attention weights & $\mathbf{W}_q^2,\mathbf{W}_k^2,\mathbf{W}_v^2$ & Vision attention weights \\
$L_{\mathrm{sf}}$ & Feature sequence length & $d_{\mathrm{sf}}$ & Feature dimension \\
$\gamma$ & DropPath factor & $\bm{\alpha}$ & RF feature extractor parameters \\
$\bm{\beta}$ & Vision feature extractor parameters & $\bm{\gamma}$ & Feature fusion module parameters \\
$\bm{\delta}$ & Semantic encoder parameters & $\bm{\epsilon}$ & Semantic decoder parameters \\
$\bm{\zeta}$ & Aggregation network parameters & $\bm{\eta}$ & Multihead prediction network parameters \\
$I_{k,n}$ & True class & $\hat{I}_{k,n}$ & Predicted class \\
$\hat{d}_{k,n}$ & Predicted distance & $\hat{\phi}_{k,n}$ & Predicted azimuth \\
$\hat{\theta}_{k,n}$ & Predicted pitch & $\hat{v}_{k,n}$ & Predicted velocity \\
$\mathcal{L}_{\mathrm{local}}$ & Local loss & $\mathcal{L}_{\mathrm{agg}}$ & Aggregation loss \\
$\mathcal{L}_d^{\mathrm{local}}$ & Local distance loss & $\mathcal{L}_d$ & Aggregation distance loss \\
$\mathcal{L}_a^{\mathrm{local}}$ & Local azimuth loss & $\mathcal{L}_a$ & Aggregation azimuth loss \\
$\mathcal{L}_p^{\mathrm{local}}$ & Local pitch loss & $\mathcal{L}_p$ & Aggregation pitch loss \\
$\mathcal{L}_v^{\mathrm{local}}$ & Local velocity loss & $\mathcal{L}_v$ & Aggregation velocity loss \\
$\mathcal{L}_i^{\mathrm{local}}$ & Local class loss & $\mathcal{L}_i$ & Aggregation class loss \\
$\Delta W = A \cdot B$ & LoRA update matrix & $A$ & LoRA matrix A \\
$B$ & LoRA matrix B & $r$ & LoRA rank \\
\bottomrule
\end{tabular}
\end{table*}

\section{Related Works}
This section reviews the related works about unimodal and multimodal sensing, as well as distributed collaborative sensing. We summarize the differences between the existing works and ours in Table \ref{tab:compare}.

\renewcommand{\arraystretch}{1.5} 
\begin{table*}[htbp]
\centering
\caption{Comparison of Our Contributions with Related Literature}
\label{tab:compare}
\renewcommand{\arraystretch}{1.2}
\begin{tabular}{lccccccccccccc}
\toprule
\textbf{Contributions} & \textbf{Ours} & \cite{9427098} & \cite{10881536} & \cite{peng2025semantic} & \cite{luo2023computer} & \cite{10478941} & \cite{peng2025simac} & \cite{9277535} & \cite{10501925} & \cite{9950719} & \cite{10242377} & \cite{9328824} & \cite{10791454} \\
\midrule
Unimodal sensing                       & \checkmark & \checkmark & \checkmark & \checkmark & \checkmark &           &           &           &           & \checkmark & \checkmark & \checkmark & \checkmark \\
Multimodal sensing                     & \checkmark &           &           &           &           & \checkmark & \checkmark & \checkmark & \checkmark &           &           &           &           \\
Distributed collaborative sensing      & \checkmark &           &           &           &           &           &           &           &           & \checkmark & \checkmark & \checkmark & \checkmark \\
Semantic-based information interaction & \checkmark &           &           & \checkmark &           &           & \checkmark &           & \checkmark &           &           &           &           \\
LLMs-based adaptive semantic decoding  & \checkmark &           &           &           &           &           &           &           &           &           &           &           &           \\
Adaptive information aggregation       & \checkmark &           &           &           &           &           &           &           &           &           &           &           &           \\
\bottomrule
\end{tabular}
\end{table*}

\subsection{Unimodal Sensing}  
Sensing technologies have demonstrated significant utility across diverse domains owing to their efficacy in target detection and tracking, particularly in complex operational environments. For example, 
Hu \textit{et al.} \cite{9427098} proposed a deep reinforcement learning-based approach to jointly optimize intelligent metasurface beamforming and signal mapping, enhancing 3D object detection accuracy in challenging RF environments.
Arslan \textit{et al.} \cite{10881536} implemented a software-defined radio (SDR) respiratory monitoring system that enhances clinical diagnostics for conditions such as pertussis, preserving patient privacy and reducing healthcare infrastructure demands. Peng \textit{et al.} \cite{peng2025semantic} proposed a computer vision (CV)-enabled edge video analysis technique that dynamically optimizes data transmission parameters through content recognition. Luo \textit{et al.} \cite{luo2023computer} established a CV-based structural health monitoring system for bridge integrity assessment through surface defect detection.  

While these implementations exploit modality-specific advantages, their reliance on unimodal data acquisition fundamentally constrains sensing diversity and situational awareness. \textit{To address these constraints, our proposed RVFN methodology synergistically integrates RF signals with visual data streams: RF modality provides robust target localization in visually degraded environments, while visual modality enhances motion parameter estimation precision through complementary sensing modalities.}

\subsection{Multimodal Sensing}  
Multimodal sensing architectures have emerged to mitigate the inherent limitations of unimodal systems, particularly regarding robustness and accuracy in complex operational scenarios. Particularly, Xu \textit{et al.} \cite{10478941} developed FVMNet, a radar-camera fusion network enabling real-time volumetric perception with demonstrated zero-shot generalization and weather-agnostic performance.
Peng \textit{et al.} \cite{peng2025simac} presented a semantic-driven integrated multimodal sensing and communication (SIMAC) framework, which introduced a multimodal semantic fusion network and an LLM-based SC model, achieving diverse sensing services and higher accuracy.
Yang \textit{et al.} \cite{9277535} developed a deep multimodal learning (DML) framework for wireless communications, introducing novel architectures to effectively fuse multi-source sensing data and demonstrating its potential through improved massive multiple-input multiple-output (MIMO) channel prediction performance.
Wang \textit{et al.} \cite{10501925} formulated a semantic simultaneous localization and mapping (SLLM) algorithm incorporating radar-visual fusion for enhanced ego-motion estimation and environment mapping.  

Despite these advancements, existing multimodal approaches primarily rely on single-device architectures with limited perspective diversity, resulting in performance degradation under occlusion scenarios. \textit{To address this limitation, our proposed TRAM incorporates a cross-aggregation attention mechanism, which enables the aggregation center to extract valuable information from collaborative devices, thus enhancing sensing accuracy.}

\subsection{Distributed Collaborative Sensing}  
Distributed collaborative architectures overcome single-view limitations by leveraging distributed sensing resources. Zhao \textit{et al.} \cite{9950719} implemented a compressed sensing framework for mechanical vibration monitoring that optimizes resource-constrained wireless sensor network (WSN) operations. Zhou \textit{et al.} \cite{10242377} devised a cloud-edge-terminal collaborative system addressing class-imbalance challenges in device-free human activity recognition. Gao \textit{et al.} \cite{9328824} introduced federated sensing, enabling elastic edge-cloud learning from decentralized sensor data. Lin \textit{et al.} \cite{10791454} developed an edge-assisted reinforcement learning (RL) scheme for connected autonomous vehicles (CAVs) perception resilience against light detection and ranging (LiDAR) interference.  

Although these studies present several innovative collaborative design frameworks, they frequently overlook the significant communication overhead arising from extensive data transmission. \textit{However, our proposed LSTN framework enables the devices to perform data transmission via the SC, reducing the communication overheads. Furthermore, LSTN harnesses the advanced capabilities of LLMs to perform sophisticated semantic decoding while incorporating physical channel information, ensuring reliable data transmission across varying channel conditions.}

\section{Distributed Collaborative Wireless Sensing and Communication System Model}
\begin{figure}[htbp]
	\centering
	\includegraphics[width=9cm]{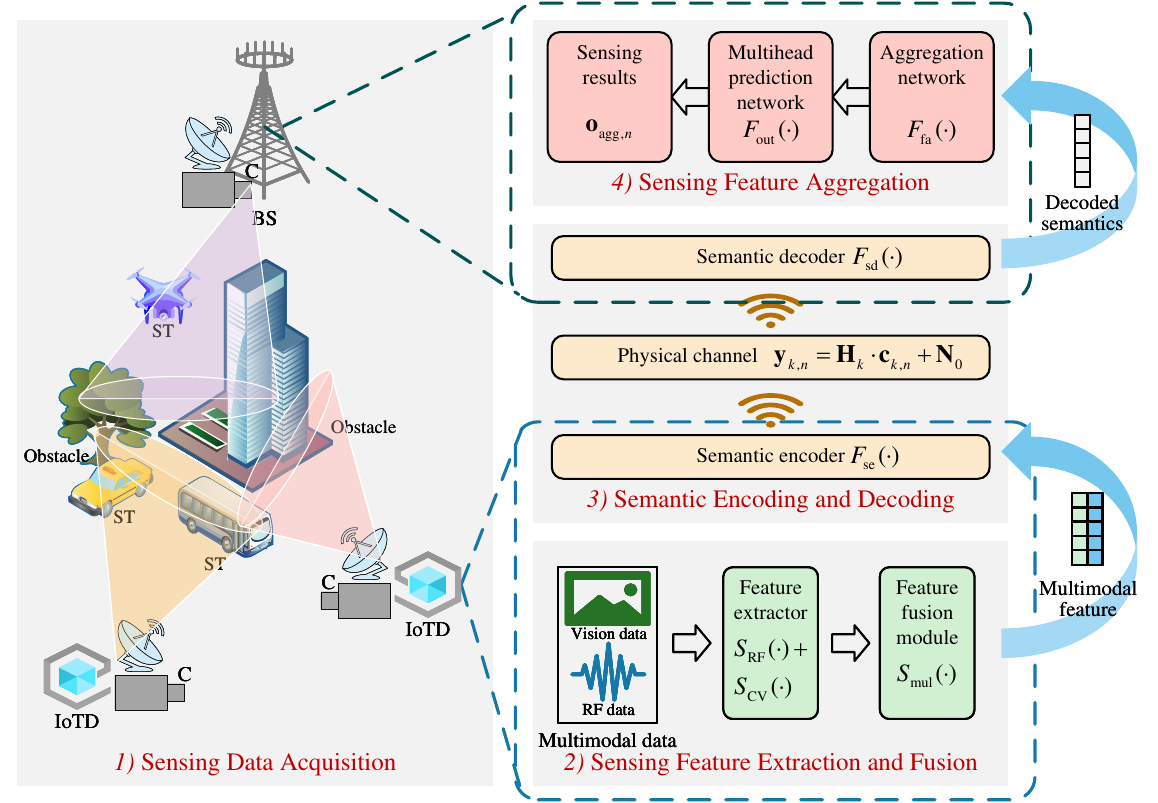}
	\caption{The illustration of the distributed collaborative wireless sensing and communication system model.}
	\label{fig:sys}
\end{figure}
\subsection{System Model}
As illustrated in Fig. \ref{fig:sys}, we consider a system consisting of multiple sensing devices, including a base station (BS), referred to as the aggregation center, $K$ Internet of Things devices (IoTDs), termed collaborative devices, and $N$ sensing targets (STs). There are some obstacles, such as buildings and trees, to hinder the sensing work.
Hence, the primary objective of these IoTDs is to support the BS by enhancing its sensing accuracy.
Each sensing device is equipped with both a camera and an RF module, enabling multimodal sensing capabilities to extract relevant features from the visual and RF data corresponding to the $n$th ST. Subsequently, each IoTD processes its local feature information and transmits it to the BS via SC. Upon receiving the data, the BS decodes the sensing features and integrates them on its own to estimate the motion parameters of the $n$th ST accurately. The detailed operational workflow is described as follows:

\subsubsection{Sensing Data Acquisition}
Assuming the $k$th IoTD serves as the reference point, the motion parameters of the $n$th ST encompass the azimuth angle \( \phi_{k,n} \), pitch angle $\theta_{k,n}$ distance \( d_{k,n} \), and radial velocity \( v_{k,n} \). We employ a single-input-multiple-output (SIMO) radar model to transmit the linear frequency modulation (LFM) waveform and capture the corresponding echo signal. The received echo signal at the $k$th IoTD can be expressed as:
\begin{equation}\label{eq:sig}
    \mathbf{R}_{k,n} = \varphi \mathbf{a}(\theta_{k,n},\phi_{k,n})\text{e}^{j2\pi\mu_{k,n}\mathbf{T}}s(\mathbf{T}-\tau_{k,n})+\mathbf{N}_0,
\end{equation}
where,
\begin{equation}
   \varphi = \frac{\lambda \cdot \epsilon_n}{(4\pi)^{3/2} \cdot d_{k,n}^2},
\end{equation}
and
\begin{equation}
    \lambda = \frac{c}{f_c},
\end{equation}
while $\epsilon_n$ denotes the radar cross-section (RCS) of the $n$th ST, which quantifies the ST's ability to reflect radar signals. $\mathbf{N}_0$ denotes the Additive White Gaussian Noise (AWGN). $s(\mathbf{T}-\tau_{k,n})=e^{j 2 \pi\left(\frac{k}{2} (\mathbf{T}-\tau_{k,n})+f_c\right) \mathbf{T}-\tau_{k,n}}$.
The steering vector is given by \( \mathbf{a}(\theta_{k,n},\phi_{k,n}) = [1, \text{e}^{-j\pi\sin\theta_{k,n}}, \ldots, \text{e}^{-j(N_z-1)\pi\sin\theta_{k,n}}]\otimes[1, \text{e}^{-j\pi\sin\phi_{k,n}\cos\theta_{k,n}}, \ldots, \text{e}^{-j(N_y-1)\pi\sin\phi_{k,n}\cos\theta_{k,n}}] \), where \( N_z \) and  \( N_y \) represents the length and width of the receiving antenna array size. The time sequence of the sampling process is denoted by \( \mathbf{T} = [k \cdot \Delta t \mid k \in \mathbb{Z}, 0 \leq k \Delta t \leq T_r ] \), where \( T_r \) is the pulse repetition interval (PRI), \( \Delta t = \frac{1}{F_s} \) is the sampling interval, and \( F_s \) is the sampling frequency. The time delay and Doppler frequency shift are represented by \( \tau_{k,n} = \frac{2d_{k,n}}{c} \) and \( \mu_{k,n} = \frac{2(f_c) v_{k,n}}{c} \), respectively, where \( f_c \) is the radar's central frequency, $K_t$ is the frequency modulation slope, and \( c = 3 \times 10^8 \, \text{m/s} \) is the speed of light.

We assume that the $k$th IoTD is equipped with a camera operating in BGR mode to capture high-quality images $\mathbf{V}_k \in \mathbb{R}^{W \times H \times 3}$, where $W$ and $H$ denote the width and height of the image in terms of the number of pixels, respectively. It is important to note that the captured image $\mathbf{V}_k$ may contain multiple STs. Therefore, our objective is to isolate the portion of the visual information corresponding to the $n$th ST, utilizing the latent information derived from the echo signal $\mathbf{R}_{k,n}$.

\subsubsection{Sensing Feature Extraction and Fusion}
Given the distinct dimensions and characteristics of the echo signal $\mathbf{R}_{k,n}$ and the captured image $\mathbf{V}_k$, we employ two separate feature extractors tailored to each modality. The process of sensing feature extraction on the $k$th IoTD is formulated as follows:
\begin{equation}
    \mathbf{s}_{k,n}^\text{RF}=S_\text{RF}(\mathbf{R}_{k,n}, \bm{\alpha}),
\end{equation}
\begin{equation}
    \mathbf{s}_{k,n}^\text{CV}=S_\text{CV}(\mathbf{V}_k, \bm{\beta}),
\end{equation}
where $\mathbf{s}_{k,n}^\text{RF}$ and $\mathbf{s}_{k,n}^\text{CV}$ denote the extracted sensing features of length $L_s$ from $\mathbf{R}_{k,n}$ and $\mathbf{V}_k$, respectively. $S_\text{RF}(\cdot)$ represents the RF feature extractor parameterized by $\bm{\alpha}$, and $S_\text{CV}(\cdot)$ corresponds to the vision feature extractor parameterized by $\bm{\beta}$. 

To effectively capture latent feature and relationships between the multimodal sensing features $\mathbf{s}_{k,n}^\text{RF}$ and $\mathbf{s}_{k,n}^\text{CV}$, a sensing feature fusion module is utilized to integrate these modalities and produce a comprehensive multimodal feature representation $\mathbf{s}_{k,n}^\text{mul}$. This fusion process is expressed as:
\begin{equation}
    \mathbf{s}_{k,n}^\text{mul}=S_\text{mul}(\mathbf{s}_{k,n}^\text{RF}, \mathbf{s}_{k,n}^\text{CV}, \bm{\gamma}),
\end{equation}
where $S_\text{mul}(\cdot)$ denotes the feature fusion module parameterized by $\bm{\gamma}$.

\subsubsection{Semantic Encoding and Decoding}
To reduce the communication overheads during wireless transmission, a semantic encoder is utilized to perform semantic encoding, with $\mathbf{s}_{k,n}^\text{mul}$ as the input. The encoding process is defined as:
\begin{equation}\label{eq:se}
    \mathbf{e}_{k,n} = F_\text{se}(\mathbf{s}_{k,n}^\text{mul},\bm{\delta}),
\end{equation}
where $\mathbf{e}_{k,n}$ denotes the semantic encoding, and $F_\text{se}(\cdot)$ represents the semantic encoder parameterized by $\bm{\delta}$. 

To enable the transmission of the semantic encoding over the wireless channel, signal modulation is employed to transform $\mathbf{e}_{k,n}$ into complex-valued symbols $\mathbf{c}_{k,n}$. These symbols are subsequently transmitted over the channel:
\begin{equation}\label{eq:trans}
	\mathbf{y}_{k,n} = \mathbf{H}_k \cdot \mathbf{c}_{k,n} + \mathbf{N}_0,
\end{equation}
where $\mathbf{y}_{k,n}$ represents the received complex-valued symbols, $\mathbf{H}_k$ denotes the channel gain between the $k$th IoTD and the BS. Given the joint source-channel coding (JSCC) architecture under consideration, the channel model must support backpropagation to facilitate end-to-end training of both the encoder and the decoder. Therefore, the wireless channel is simulated using neural network-based methodologies \cite{jiang2024large}. The transmission rate is expressed as:
\begin{equation}
	V_k = B_k \log _{2} \left( 1 + \frac{P_k \mathbf{H}_k}{\mathbf{N}_0} \right),
\end{equation}
where $B_k$ and $P_k$ denote the allocated bandwidth and transmission power. Consequently, the transmission delay is given by:
\begin{equation}
	t^\text{com}_k = \frac{Z(\mathbf{c}_{k,n})}{V_k},
\end{equation}
where $Z(\mathbf{c}_{k,n})$ represents the number of bits required to transmit the complex-valued symbols $\mathbf{c}_{k,n}$.

Upon receiving the symbols $\mathbf{y}_{k,n}$, signal demodulation is applied to reconstruct the received semantic encoding $\hat{\mathbf{e}}_{k,n}$. A semantic decoder is then employed to recover the sensing feature $\mathbf{\hat{s}}_{k,n}^\text{mul}$, which is formulated as:
\begin{equation}\label{eq:sd}
    \mathbf{\hat{s}}_{k,n}^\text{mul} = F_\text{sd}(\hat{\mathbf{e}}_{k,n}, \bm{\epsilon}),
\end{equation}
where $F_\text{sd}(\cdot)$ denotes the semantic decoder parameterized by $\bm{\epsilon}$.

\subsubsection{Sensing Feature Aggregation}
Upon receiving all the sensing features $\mathbf{\hat{s}}_{n}^\text{mul} = \{\mathbf{\hat{s}}_{\text{agg},n}^\text{mul}, \mathbf{\hat{s}}_{1,n}^\text{mul},..., \mathbf{\hat{s}}_{K,n}^\text{mul}\}$, where $\mathbf{\hat{s}}_{\text{agg},n}^\text{mul}$ represents the feature information of the aggregation center. Thereafter, an aggregation network is employed to integrate them, generating the aggregated feature:
\begin{equation}
    \textbf{S}_{n}^\text{agg} = F_\text{fa}(\mathbf{\hat{s}}_{n}^\text{mul}, \bm{\zeta}),
\end{equation}
where $F_\text{fa}(\cdot)$ denotes the aggregation network parameterized by $\bm{\zeta}$. To meet the diverse sensing requirements, a multihead prediction network is used to predict the sensing results:
\begin{equation}
    \textbf{o}_{\text{agg},n} = F_\text{out}(\mathbf{S}_{n}^\text{agg}, \bm{\eta}),
\end{equation}
where $F_\text{out}(\cdot)$ is the multihead prediction network parameterized by $\bm{\eta}$. The sensing results, $\textbf{o}_{\text{agg},n} \in \{\hat{\phi}_{\text{agg},n}, \hat{\theta}_{\text{agg},n}, \hat{d}_{\text{agg},n}, \hat{v}_{\text{agg},n}, \mathbf{\hat{I}}_{\text{agg},n}\}$, encompass the predicted distance $\hat{d}_{\text{agg},n}$, radial velocity $\hat{v}_{\text{agg},n}$, azmuith angle $\hat{\phi}_{\text{agg},n}$, pitch angle $\hat{\theta}_{\text{agg},n}$, and the class of the $n$th ST $\mathbf{\hat{I}}_{\text{agg},n}$.

\subsection{Problem Formulation}
When performing the distributed collaborative sensing, the total execution time $T_k^{\mathrm{exe}}$ for the $k$th IoTD comprises the computation time for feature extraction $t_k^{\mathrm{ft}}$, semantic encoding $t_k^{\mathrm{se}}$, the communication time for transmission $t_k^{\mathrm{com}}$, the computation time for semantic decoding $t_k^{\mathrm{sd}}$, and the computation time for feature aggregation $t_k^{\mathrm{fa}}$. Thus, the total execution time is formulated as:
\begin{equation}
T^{\mathrm{exe}} = \max(\{t_k^{\mathrm{ft}} + t_k^{\mathrm{se}} + t_k^{\mathrm{com}} + t_k^{\mathrm{sd}} + t_k^{\mathrm{fa}} \mid 1 \leq k \leq K \}).
\end{equation}

To accommodate diverse sensing tasks, including distance, velocity, and angle prediction, as well as the identification of the $n$th ST, the corresponding task losses are defined as follows:
\begin{equation}\label{eq:loss1}
    \mathcal{L}_\text{d} = ||d_{\text{agg},n} - \hat{d}_{\text{agg},n}||^2,
\end{equation}
\begin{equation}\label{eq:loss2}
    \mathcal{L}_\text{a} = ||\phi_{\text{agg},n} - \hat{\phi}_{\text{agg},n}||^2,
\end{equation}
\begin{equation}\label{eq:loss3}
    \mathcal{L}_\text{p} = ||\theta_{\text{agg},n} - \hat{\theta}_{\text{agg},n}||^2,
\end{equation}
\begin{equation}\label{eq:loss4}
    \mathcal{L}_\text{v} = ||v_{\text{agg},n} - \hat{v}_{\text{agg},n}||^2,
\end{equation}
\begin{equation}\label{eq:loss5}
    \mathcal{L}_\text{i} = \operatorname{CE}(\mathbf{I}_{\text{agg},n}, \mathbf{\hat{I}}_{\text{agg},n}) = -\sum_{i=1}^{M} I_{i} \log \left(\hat{I}_{i}\right),
\end{equation}
where $d_{\text{agg},n}$, $\phi_{\text{agg},n}$, $\theta_{\text{agg},n}$, $v_{\text{agg},n}$, and $I_n$ represent the distance, azimuth angle, pitch angle, radial velocity, and class of the $n$th ST, with the aggregation center (BS) as the reference point. $\mathbf{I}_{\text{agg},n} = [I_1, I_2, \dots, I_M]$ denotes the one-hot encoded labels, where $I_i = 1$ if the $n$th ST belongs to the $i$-th class and $0$ otherwise. $\mathbf{\hat{I}}_{\text{agg},n} = [\hat{I}_1, \hat{I}_2, \dots, \hat{I}_M]$ represents the predicted probabilities, with $\hat{I}_i$ indicating the probability of the $n$th ST being classified as the $i$-th class. $M$ denotes the total number of categories. 

The primary objective of the LLM-DiSAC framework is to minimize semantic distortion during wireless transmission, thus maximizing the accuracy of the predicted sensing results. Furthermore, execution delays must adhere to the quality of service requirements. Consequently, the objective function of the proposed LLM-DiSAC framework is formulated as:  
\begin{subequations}\label{eq:problem}  
\begin{align}  
\min_{\bm{\alpha}, \bm{\beta}, \bm{\gamma}, \bm{\delta}, \bm{\epsilon}, \bm{\zeta}} l_1\mathcal{L}_\text{d} + l_2\mathcal{L}_\text{a} + l_3\mathcal{L}_\text{p} + l_4\mathcal{L}_\text{v} + l_5\mathcal{L}_\text{i},  
\end{align}  
\begin{alignat}{1}  
\text{s.t.~}  
& T^\text{exe} \leq T_\text{max},  
\end{alignat}  
\end{subequations}  
where $T_\text{max}$ represents the maximum allowable latency for completing the sensing task, and $l_1$, $l_2$, $l_3$, $l_4$, and $l_5$ are weighting factors that adjust the relative importance of each loss component.

To address the optimization problem Eq. (\ref{eq:problem}a), three critical challenges must be resolved. First, vision and RF signals exhibit distinct modalities, making it difficult to process them effectively using a single neural network.  
Second, traditional aggregation methods rely on limited-capacity discriminative AI models and fail to incorporate physical channel information, leading to inaccurate sensing accuracy.  
Finally, frequent data transmission between the BS and IoTDs incurs significant communication overhead, which cannot be overlooked.  
Therefore, we have meticulously designed specialized neural networks for each module within the LLM-DiSAC framework. The details of these designs will be elaborated in the following section.

\section{LLM-Driven Distributed Integrated Multimodal Sensing and SC}
\begin{figure*}[htbp]
	\centering
	\includegraphics[width=16cm]{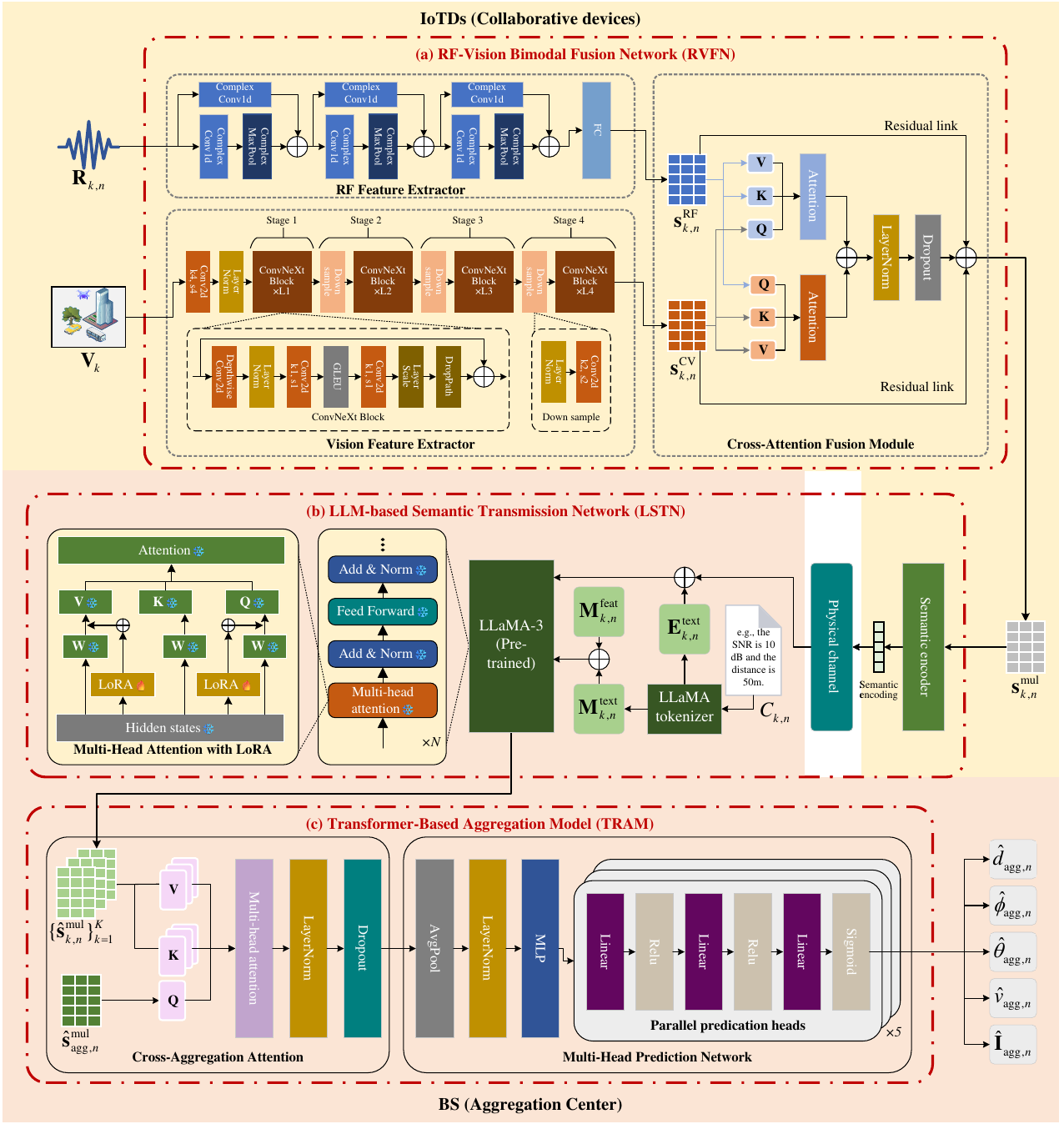}
	\caption{The illustration of the LLM-DiSAC framework.}
	\label{fig:LLM-DiSAC}
\end{figure*}
\subsection{Overview}
Conventional sensing systems and lightweight AI models struggle with limited awareness, fixed perspectives, and high communication costs, hindering efficient edge sensing. To address this, we propose LLM-DiSAC—a novel framework that integrates multimodal, multi-device sensing with SC for accurate and low-cost deployment. As illustrated in Fig. \ref{fig:LLM-DiSAC}, this framework comprises the following core modules:

\subsubsection{RVFN for Multimodal Feature Representation}
Given the heterogeneity of multimodal data, RVFN leverages complex CNNs and hybrid visual networks as feature extractors, \( S_\text{RF}(\cdot) \) and \( S_\text{CV}(\cdot) \), for processing RF and visual modalities, respectively. A fusion network \( S_\text{mul}(\cdot) \), based on a cross-attention mechanism, is then employed to perform deep feature fusion, resulting in a multimodal feature representation, \( \mathbf{s}_{k,n}^\text{mul} \). (Details in Section IV-B.)
Theoretically, the multimodal feature representation encapsulates the critical information inherent in the original multimodal data, making it suitable as the source input for SC.  

\subsubsection{LSTN for Channel-Adaptive and Low-cost Transmission}
Traditional communication systems struggle with dynamic channel conditions (e.g., varying SNR and transceiver distances). To address this, the LSTN leverages an LLM-based semantic decoder that adapts to real-time communication parameters. Thus, Eq. (\ref{eq:sd}) is reformulated as:  
\begin{equation}
    \mathbf{\hat{s}}_{k,n}^\text{mul} = F_\text{sd}(\hat{\mathbf{e}}_{k,n},C_{k,n}, \bm{\epsilon}),
\end{equation}
where $C_{k,n}$ represents the communication parameters that are in the format of natural languages (e.g., ``the SNR is 5 dB and the distance is 50m"). (Details in Section IV-C.)
By dynamically adjusting \( C_{k,n} \), the semantic decoder ensures robust performance across diverse channel conditions. The decoded features $\mathbf{\hat{s}}_{k,n}^\text{mul}$ are then aggregated to optimize sensing accuracy.  

\subsubsection{TRAM for Multidevice Information Aggregation}
Due to the limited field of view, a single device is often unable to effectively perceive specific obscured STs. To address this limitation, the TRAM allows individual sensing devices to transmit their local multi-modal sensing features, $\mathbf{\hat{s}}_{n}^\text{mul}$, to an aggregation center. An aggregation attention mechanism is then employed to extract the most relevant information from these features, enhancing the overall sensing performance by fusing them with the center’s feature. Finally, a multi-head prediction network generates task-specific sensing results $\mathbf{o}_{\text{agg},n}$ based on the aggregated features $\mathbf{S}_n^\text{agg}$. (Details in Section IV-D.)
This approach effectively mitigates the perceptual limitations of individual devices caused by constrained viewing angles.

\subsubsection{Privacy-Aware Distributed Training}
To ensure data privacy, we propose a novel two-stage distributed learning strategy. In the first stage, each sensing device independently trains the RVFN and LSTN models using local data. In the second stage, only the intermediate feature representation $\mathbf{e}_{k,n}$ is transmitted to the aggregation center, where the TRAM model is subsequently trained. (Details in Section IV-E.)

Assuming $\mathcal{D}$ denotes the training dataset, the overall workflow of the proposed LLM-DiSAC framework is presented in \textbf{Algorithm \ref{alg:LLM-DiSAC}}. 
Different from traditional sensing systems, LLM-DiSAC employs a multimodal sensing approach to improve the accuracy and stability of sensing. 
Furthermore, a distributed collaboration mechanism is introduced to enhance the sensing performance of individual devices while leveraging the capabilities of LLMs to achieve highly robust SC. This innovative design effectively addresses the limitations of conventional methods, significantly reduces communication overhead, and substantially improves noise resilience.

\begin{algorithm}[htbp]
    \caption{LLM-DiSAC Framework Workflow}
    \label{alg:LLM-DiSAC}
    \begin{algorithmic}[1]
        \REQUIRE \(\mathcal{D}\).
        \ENSURE \(\mathbf{o}_{\text{agg},n}, \bm{\alpha}, \bm{\beta}, \bm{\gamma}, \bm{\delta}, \bm{\epsilon}, \bm{\zeta},\bm{\eta}.\)  

        \vspace{0.5em}
        \noindent\textbf{Inference Phase:}
        \FOR{each sensing device}
        \STATE Obtain the semantic representation \(\mathbf{s}_{k,n}^\text{mul}\) from multimodal data \(\mathbf{V}_k\) and \(\mathbf{R}_{k,n}\) using \textbf{Algorithm \ref{alg:RVFN}}.
        \STATE Perform wireless communication and obtain the channel-adaptive semantic decoding $\mathbf{\hat{s}}_{k,n}^\text{mul}$ using \textbf{Algorithm \ref{alg:LSTN}}.
        \ENDFOR
        \STATE{Based on all sensing features $\mathbf{\hat{s}}_{n}^\text{mul}$, predict the sensing results $\mathbf{o}_{\text{agg},n}$ using \textbf{Algorithm \ref{alg:TRAM}}.}

        \vspace{0.5em}
        \noindent\textbf{Training Phase:}
        \STATE Obtain the trained parameters: \(\bm{\alpha}, \bm{\beta}, \bm{\gamma}, \bm{\delta}, \bm{\epsilon},\bm{\zeta},\bm{\eta},\) by training all the modules according to \textbf{Algorithm \ref{alg:TDL}}, using \(\mathcal{D}\).
    \end{algorithmic}
\end{algorithm}

\subsection{RF-Vision Bimodal Fusion Network}
The RVFN adopts a hybrid architecture for feature extraction: a ConvNet-based visual branch and a complex-valued RF branch. These unimodal features are integrated via a bidirectional cross-attention mechanism to produce unified multimodal representations, as shown in Fig. \ref{fig:LLM-DiSAC} (a). Key modules are detailed below:

\subsubsection{RF Feature Extractor}
The RF signal \(\mathbf{R}_{k,n} \in \mathbb{C}^{B \times T \times L_\text{RF}}\) is processed using complex-valued convolutional layers, each operating separately on real and imaginary parts:
\begin{equation}\label{eq:RFE1}
    \mathbf{z}_{k,n}^\text{real} = \mathbf{W}_\text{real} \ast \mathbf{x}^\text{real} - \mathbf{W}_\text{imag} \ast \mathbf{x}^\text{imag},    
\end{equation}
\begin{equation}\label{eq:RFE2}
    \mathbf{z}_{k,n}^\text{imag} = \mathbf{W}_\text{real} \ast \mathbf{x}^\text{imag} + \mathbf{W}_\text{imag} \ast \mathbf{x}^\text{real},
\end{equation}
\begin{equation}\label{eq:RFE3}
    \mathbf{z}_{k,n}^\text{out} = \text{ReLU}(\mathbf{z}_{k,n}^\text{real} + j \cdot \mathbf{z}_{k,n}^\text{imag}).
\end{equation}

To hierarchically extract features, complex max-pooling follows each convolution. After three stages, the final outputs are concatenated:
\begin{equation}\label{eq:RFE4}
    \mathbf{z}_{k,n,3} = \text{Concat}(\mathbf{z}_{k,n,3}^\text{real}, \mathbf{z}_{k,n,3}^\text{imag}).
\end{equation}

To a reduced dimensionality, a fully connected layer maps the concatenated features, and we obtain the RF feature:
\begin{equation}\label{eq:RFE5}
    \mathbf{s}_{k,n}^\text{RF} = \text{Linear}(\mathbf{z}_{k,n,3}), \mathbf{s}_{k,n}^\text{RF} \in \mathbb{R}^{B \times L_\text{sf} \times d_\text{sf}}.
\end{equation}

\subsubsection{Vision Feature Extractor}
To ensure high extraction accuracy and fast inference, we utilize the ConvNet \cite{Liu_2022_CVPR} to balance performance and speed, combining ViT design elements with convolutional efficiency. The input image \(\mathbf{V}_k \in \mathbb{R}^{3 \times H \times W}\) is first embedded via a \(4 \times 4\) convolution:
\begin{equation}\label{eq:VE1}
\mathbf{f}_1 = \text{Conv}(\mathbf{V}_k) \in \mathbb{R}^{C_1 \times H/4 \times W/4}.
\end{equation}

The network has four stages, each with stacked ConvNeXt blocks and optional downsampling:
\begin{equation}\label{eq:VE2}
\mathbf{f}_{i+1} = 
\begin{cases}
\text{DS}_i(\mathcal{B}_i(\mathbf{f}_i)), & \text{if downsampling},\\
\mathcal{B}_i(\mathbf{f}_i), & \text{otherwise},
\end{cases}
\end{equation}
where \(\text{DS}_i\) is a \(2 \times 2\) convolution, and \(\mathcal{B}_i\) includes \(L_i\) ConvNeXt blocks. Each block applies depthwise convolution and MLP:
\begin{equation}\label{eq:VE3}
\mathbf{z}_1 = \text{LayerNorm}(\text{DWConv}(\mathbf{z})), \end{equation}
\begin{equation}\label{eq:VE4}
\mathbf{z}_2 = \mathbf{W}_2 \cdot \text{GLEU}(\mathbf{W}_1 \cdot \mathbf{z}_1),
\end{equation}
followed by a residual connection with stochastic depth:
\begin{equation}\label{eq:VE5}
    \mathbf{s}_{k,n}^\text{CV} = \mathbf{z} + \gamma \cdot \text{DropPath}(\mathbf{z}_2), \mathbf{s}_{k,n}^\text{CV} \in \mathbb{R}^{B \times L_\text{sf} \times d_\text{sf}}.
\end{equation}

By stacking multiple such blocks in each stage, the network progressively transforms low-level features into high-level semantic representations while maintaining stable gradient flow.

\subsubsection{Cross-Attention Fusion Module}
To achieve the deep multimodal fusion, we integrate the RF feature $\mathbf{s}_{k,n}^\text{RF}$ and vision feature $\mathbf{s}_{k,n}^\text{CV}$ through a bidirectional attention mechanism \cite{liu2019bidirectional}. Specifically, $\mathbf{s}_{k,n}^\text{RF}$ acts as the query, while $\mathbf{s}_{k,n}^\text{CV}$ serves as the key and value, and vice versa. The attention outputs are computed as:
\begin{equation}\label{eq:crt1}
    \mathbf{z}_{k,n}^\text{CV} = \text{Attention}(\mathbf{W}_q^1 \mathbf{s}_{k,n}^\text{RF}, \mathbf{W}_k^2 \mathbf{s}_{k,n}^\text{CV}, \mathbf{W}_v^2 \mathbf{s}_{k,n}^\text{CV}),
\end{equation}
\begin{equation}\label{eq:crt2}
    \mathbf{z}_{k,n}^\text{RF} = \text{Attention}(\mathbf{W}_q^2 \mathbf{s}_{k,n}^\text{CV}, \mathbf{W}_k^1 \mathbf{s}_{k,n}^\text{RF}, \mathbf{W}_v^1 \mathbf{s}_{k,n}^\text{RF}),
\end{equation}
where \(\mathbf{W}_q^1, \mathbf{W}_k^1, \mathbf{W}_v^1\) are the query, key, and value weights for radar signals, and \(\mathbf{W}_q^2, \mathbf{W}_k^2, \mathbf{W}_v^2\) are the corresponding weights for image features. The fused output is the sum of the two attention outputs:
\begin{equation}\label{eq:crt3}
\mathbf{z}_{k,n}^\text{fusion} = \mathbf{z}_{k,n}^\text{CV} + \mathbf{z}_{k,n}^\text{RF}.
\end{equation}

To avoid the issue of vanishing gradient, normalization and residual connections are employed to refine the fusion:
\begin{equation}\label{eq:crt4}
\mathbf{s}_{k,n}^\text{mul} = \text{LayerNorm}(\mathbf{z}_{k,n}^\text{fusion}) + \mathbf{s}_{k,n}^\text{RF} + \mathbf{s}_{k,n}^\text{CV}, \mathbf{s}_{k,n}^\text{mul} \in \mathbb{R}^{B \times L_\text{sf} \times d_\text{sf}}.
\end{equation}

Overall, RVFN effectively combines complex-valued RF processing, convolutional vision modeling, and attention-based fusion, enabling robust bimodal feature learning. The inference procedure is outlined in \textbf{Algorithm \ref{alg:RVFN}}.

\begin{algorithm}[htbp]
\caption{Inference of RVFN}
\label{alg:RVFN}
\begin{algorithmic}[1]
	\REQUIRE $\mathbf{V}_k$, $\mathbf{R}_{k,n}$.
	\ENSURE $\mathbf{s}_{k,n}^\text{mul}$.
	\STATE{Extract RF feature $\mathbf{s}_{k,n}^\text{RF}$ using Eqs. (\ref{eq:RFE1})-(\ref{eq:RFE5}).}
	\STATE{Extract vision feature $\mathbf{s}_{k,n}^\text{CV}$ using Eqs. (\ref{eq:VE1})-(\ref{eq:VE5}).}
	\STATE{Obtain the fused multimodal feature $\mathbf{s}_{k,n}^\text{mul}$ via cross-attention using Eqs. (\ref{eq:crt1})-(\ref{eq:crt4}).}	
\end{algorithmic}
\end{algorithm}

\subsection{LLM-based Semantic Transmission Network}

To reduce the communication overhead associated with transmitting multimodal features from collaborative devices to the aggregation center, while simultaneously enhancing robustness against channel noise, the LSTN is introduced to perform both semantic encoding and decoding. As shown in Fig. \ref{fig:LLM-DiSAC} (b), the detailed workflow is described as follows.

First, given the input multimodal feature \( \mathbf{s}_{k,n}^\text{mul} \) (extracted via the RVFN), the semantic encoder applies a linear transformation to project the input into a lower-dimensional semantic space suitable for efficient transmission:
\begin{equation}\label{eq:LSTN1}
  \mathbf{e}_{k,n} = \text{Linear}(\mathbf{s}_{k,n}^\text{mul}), \quad \mathbf{e}_{k,n} \in \mathbb{R}^{B \times L_\text{se} \times d_{\text{se}}},
\end{equation}
where \( L_\text{se} \) denotes the sequence length after semantic encoding and \( d_\text{se} \) is the reduced feature dimensionality. Note that \( L_\text{se} \ll L_\text{sf} \) and \( d_\text{se} \ll d_\text{sf} \), implying a significant reduction in the size of the data to be transmitted.

Second, after transmission through a noisy wireless channel, the received semantic representation \( \hat{\mathbf{e}}_{k,n} \), along with the corresponding textual communication parameters \( C_{k,n} \), are input into the semantic decoder. The decoder is built upon the LLaMA-3 model \cite{grattafiori2024llama}, which is in turn built on the transformer architecture and enhanced with low-rank adaptation (LoRA) \cite{hu2022lora} for efficient fine-tuning in downstream tasks. 
The textual communication parameters $C_{k,n}$ are tokenized and mapped into the embedding space:
\begin{equation}\label{eq:LSTN2}
    \mathbf{E}_{k,n}^\text{text} = \text{Embed}(C_{k,n}), \quad \mathbf{E}_{k,n}^\text{text} \in \mathbb{R}^{B \times L_\text{text} \times d_\text{sd}},
\end{equation}
where \( L_\text{text} \) is the tokenized sequence length and \( d_\text{sd} \) is the embedding dimension of the decoder.

The received semantic encoding \( \hat{\mathbf{e}}_{k,n} \) and the textual embeddings \( \mathbf{E}_{k,n}^\text{text} \) are then concatenated along the sequence dimension to form a unified multimodal input:
\begin{equation}\label{eq:LSTN3}
    \mathbf{E}_{k,n} = \text{Concat}(\hat{\mathbf{e}}_{k,n}, \mathbf{E}_{k,n}^\text{text}), \quad \mathbf{E}_{k,n} \in \mathbb{R}^{B \times L_\text{fusion} \times d_\text{sd}},
\end{equation}
where \( L_\text{fusion} = L_\text{se} + L_\text{text} \) denotes the total input sequence length.

Simultaneously, an attention mask is generated to guide the attention mechanism within the transformer decoder. 
The textual input $C_{k,n}$ undergoes tokenization using the LLaMA-3 tokenizer, producing the corresponding attention mask \(\mathbf{M}_{k,n}^\text{text}\). Therefore, the attention mask is constructed by concatenating a visual feature mask \( \mathbf{M}_{k,n}^\text{feat} \) (initialized as an all-one matrix) and a textual attention mask \( \mathbf{M}_{k,n}^\text{text} \):
\begin{equation}\label{eq:LSTN4}
    \mathbf{F}_{k,n}^\text{mask} = \text{Concat}(\mathbf{M}_{k,n}^\text{feat}, \mathbf{M}_{k,n}^\text{text}), \quad \mathbf{F}_{k,n}^\text{mask} \in \mathbb{R}^{B \times L_\text{fusion}}.
\end{equation}

Finally, the concatenated input and attention mask are fed into the LLaMA-3 decoder, yielding the recovered multimodal sensing feature:
\begin{equation}\label{eq:LSTN5}
    \mathbf{\hat{s}}_{k,n}^\text{mul} = \text{LLaMA3}(\mathbf{E}_{k,n}, \mathbf{F}_{k,n}^\text{mask}), \mathbf{\hat{s}}_{k,n}^\text{mul} \in \mathbb{R}^{B \times L_\text{fusion} \times d_\text{sd}}.
\end{equation}

The complete inference procedure of the LSTN is summarized in \textbf{Algorithm~\ref{alg:LSTN}}. By integrating the LLaMA-3 language model into the SC system, the LSTN enables robust and accurate multimodal feature reconstruction. Moreover, the use of LoRA significantly reduces the number of trainable parameters during adaptation, facilitating lightweight and efficient deployment in communication-constrained environments. As a result, the LSTN effectively bridges the gap between textual and non-textual modalities, enhancing both the transmission efficiency and the downstream feature aggregation performance.
\begin{algorithm}
\caption{Inference of LSTN}
\label{alg:LSTN}
\begin{algorithmic}[1]
    \REQUIRE $\mathbf{s}_{k,n}^\text{mul}$, $C_{k,n}$.
    \ENSURE $\mathbf{\hat{s}}_{k,n}^\text{mul}$.
    \STATE{Tokenize $C_{k,n}$ to obtain $\mathbf{E}_{k,n}^\text{text}$ using Eq. (\ref{eq:LSTN2}).}
    \STATE{Concatenate $\hat{\mathbf{e}}_{k,n}$ and $\mathbf{E}_{k,n}^\text{text}$ to obtain $\mathbf{E}_{k,n}$ using Eq. (\ref{eq:LSTN3}).}
    \STATE{Generate the attention mask $\mathbf{F}_{k,n}^\text{mask}$ using Eq. (\ref{eq:LSTN4}).}
    \STATE{Generate recovered sensing feature $ \mathbf{\hat{s}}_{k,n}^\text{mul}$ using LLaMA-3 according to Eq. (\ref{eq:LSTN5}).}
\end{algorithmic}
\end{algorithm}

\subsection{Transformer-Based Aggregation Model}
TRAM is designed to aggregate local features from multiple collaborative devices, thereby enhancing the sensing accuracy of the aggregation center. The core innovation of TRAM lies in its cross-aggregation attention mechanism, which adaptively integrates feature information from various collaborative sensing devices and extracts the most relevant latent features. As shown in Fig. \ref{fig:LLM-DiSAC} (c), the detailed process is described as follows.

First, the aggregation attention aims to refine the main feature \( \mathbf{\hat{s}}_{\text{agg},n}^\text{mul} \) by aggregating contextual information from a set of supporting features \( \{\mathbf{\hat{s}}_{k,n}^\text{mul}\}_{k=1}^K \), using a stack of \( L \) multi-head attention layers. Each layer includes learnable query, key, and value projections, followed by standard multi-head attention computation. For each layer \( l \in \{1,\dots,L\} \), the process is formulated as:
\begin{equation}\label{eq:TRAM1}
    \mathbf{Q}_n^{(l)} = \mathbf{\hat{s}}_{\text{agg},n}^{(l)} \mathbf{W}_Q^{(l)}, \quad
    \mathbf{K}_{k,n}^{(l)} = \mathbf{\hat{s}}_{k,n}^\text{mul} \mathbf{W}_K^{(l)}, \quad
    \mathbf{V}_{k,n}^{(l)} = \mathbf{\hat{s}}_{k,n}^\text{mul} \mathbf{W}_V^{(l)},
\end{equation}
\begin{equation}\label{eq:TRAM2}
    \mathbf{F}_{k,n}^{(l)} = \text{MultiHeadAttn}(\mathbf{Q}_n^{(l)}, \mathbf{K}_{k,n}^{(l)}, \mathbf{V}_{k,n}^{(l)}),
\end{equation}
where $\mathbf{W}_Q^{(l)}$, $\mathbf{W}_K^{(l)}$, and $\mathbf{W}_V^{(l)}$ are the query, key, and value weights. The attention outputs \( \mathbf{F}_{k,n}^{(l)} \) from all collaborative devices are then aggregated—e.g., via summation—to update the main sensing feature at the aggregation center:
\begin{equation}\label{eq:TRAM3}
    \mathbf{\hat{s}}_{\text{agg},n}^{(l+1)} = \sum_{k=1}^K \mathbf{F}_{k,n}^{(l)}.
\end{equation}

After passing through all \( L \) attention layers, the final aggregated representation is normalized and regularized as:
\begin{equation}\label{eq:TRAM4}
    \mathbf{S}_n^\text{agg} = \text{Dropout}(\text{LayerNorm}(\mathbf{\hat{s}}_{\text{agg},n}^{(L)})), \mathbf{S}_n^\text{agg} \in \mathbb{R}^{B \times L_\text{fusion} \times d_\text{sd}}.
\end{equation}

Next, the aggregated feature representation \( \mathbf{S}_n^\text{agg} \) is input to a shared multi-head prediction network, which estimates the motion parameters of the \( n \)th ST. An adaptive average pooling operation followed by normalization is applied to reduce the dimension:
\begin{equation}\label{eq:TRAM5}
    \mathbf{g}_n = \text{MLP}(\text{Norm}(\text{AvgPool}(\mathbf{S}_n^\text{agg}))).
\end{equation}

The resulting vector \( \mathbf{g}_n \) is then fed into five parallel prediction heads to estimate the respective parameters:
\begin{subequations}\label{eq:TRAM6}
\begin{align}
    \hat{d}_{\text{agg},n} &= \text{Sigmoid}(\mathbf{W}_d \mathbf{g}_n), \\
    \hat{\phi}_{\text{agg},n} &= \text{Sigmoid}(\mathbf{W}_a \mathbf{g}_n), \\
    \hat{\theta}_{\text{agg},n} &= \text{Sigmoid}(\mathbf{W}_p \mathbf{g}_n), \\
    \hat{v}_{\text{agg},n} &= \text{Sigmoid}(\mathbf{W}_v \mathbf{g}_n), \\
    \mathbf{\hat{I}}_{\text{agg},n} &= \mathbf{W}_c \mathbf{g}_n,
\end{align}
\end{subequations}
where \( \mathbf{W}_d, \mathbf{W}_a, \mathbf{W}_p, \mathbf{W}_v, \) and \( \mathbf{W}_c \) are the learnable projection matrices for each respective branch.

The inference workflow of TRAM is summarized in \textbf{Algorithm~\ref{alg:TRAM}}. By leveraging its cross-aggregation attention mechanism, TRAM enables the aggregation center to adaptively fuse complementary sensing features from collaborative devices, significantly enhancing overall sensing accuracy, even under fixed-viewpoint constraints.
\begin{algorithm}
\caption{Inference of TRAM}
\label{alg:TRAM}
\begin{algorithmic}[1]
    \REQUIRE $\mathbf{\hat{s}}_{n}^\text{mul}$.
    \ENSURE $\mathbf{o}_{\text{agg},n}$.
    \STATE{Perform attention computation to $\mathbf{\hat{s}}_{n}^\text{mul}$ using Eqs. (\ref{eq:TRAM1})-(\ref{eq:TRAM2}).}
    \STATE{Perform feature aggregation and obtain $\mathbf{S}_n^\text{agg}$ using Eqs. (\ref{eq:TRAM3})-(\ref{eq:TRAM4}).}
    \STATE{Predict the sensing results $\mathbf{o}_{\text{agg},n}$ using Eqs. (\ref{eq:TRAM5})-(\ref{eq:TRAM6}).}
\end{algorithmic}
\end{algorithm}

\subsection{Two-Stage Distributed Learning}
\begin{figure}[htbp]
	\centering
	\includegraphics[width=8.5cm]{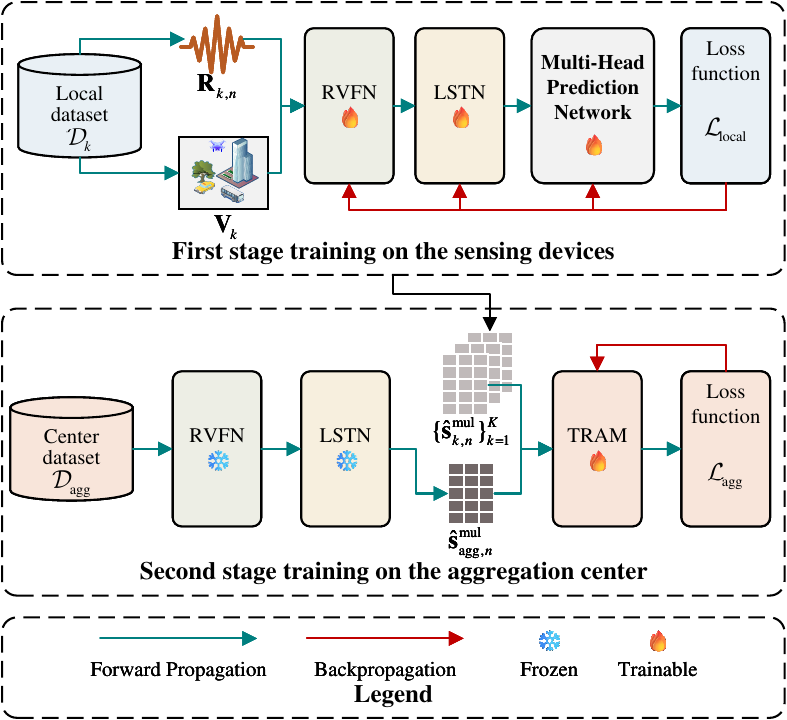}
	\caption{The illustration of the two-stage distributed learning.}
	\label{fig:TDL}
\end{figure}

We propose a novel two-stage distributed learning strategy to enable the LLM-DiSAC framework to provide diversified sensing services while preserving data privacy. In this strategy, the RVFN, LSTN, and TRAM modules are trained independently. As illustrated in Fig. \ref{fig:TDL}, the process is as follows.

In the first stage, to ensure that the features extracted by RVFN and LSTN are task-relevant, a multi-head prediction network is appended to the end of the LSTN. This enables each sensing device to optimize RVFN and LSTN via supervised learning using its own dataset $\mathcal{D}_k$. The supervised learning loss function is defined as:
\begin{equation}\label{eq:TDL1}
\mathcal{L}_\text{local} = l_1\mathcal{L}_\text{d}^\text{local} + l_2\mathcal{L}_\text{a}^\text{local} + l_3\mathcal{L}_\text{p}^\text{local} + l_4\mathcal{L}_\text{v}^\text{local} + l_5\mathcal{L}_\text{i}^\text{local},  
\end{equation}
where,
\begin{align}
    \mathcal{L}_\text{d}^\text{local} &= \|\hat{d}_{k,n} - d_{k,n}\|^2, \label{eq:TDL2} \\
    \mathcal{L}_\text{a}^\text{local} &= \|\hat{\phi}_{k,n} - \phi_{k,n}\|^2, \label{eq:TDL3} \\
    \mathcal{L}_\text{p}^\text{local} &= \|\hat{\theta}_{k,n} - \theta_{k,n}\|^2, \label{eq:TDL4} \\
    \mathcal{L}_\text{v}^\text{local} &= \|\hat{v}_{k,n} - v_{k,n}\|^2, \label{eq:TDL5} \\
    \mathcal{L}_\text{i}^\text{local} &= \text{CE}(\hat{\mathbf{I}}_{k,n}, \mathbf{I}_{k,n}), \label{eq:TDL6}
\end{align}
where $d_{k,n}$, $\phi_{k,n}$, $\theta_{k,n}$, $v_{k,n}$, and $I_{k,n}$ denote the ground truth distance, azimuth angle, pitch angle, radial velocity, and class label of the $n$th ST, with respect to the $k$th IoTD. Correspondingly, $\hat{d}_{k,n}$, $\hat{\phi}_{k,n}$, $\hat{\theta}_{k,n}$, $\hat{v}_{k,n}$, and $\hat{I}_{k,n}$ are the predicted values.

To enhance training efficiency, we fine-tune only the last four layers of the LLaMA backbone in the LSTN using LoRA \cite{hu2022lora}. LoRA is a parameter-efficient fine-tuning method that inserts low-rank decomposition matrices into the transformer layers, significantly reducing the number of trainable parameters. Specifically, instead of updating the full weight matrix \( W \), LoRA introduces a low-rank update:
\begin{equation}\label{eq:TDL7}
    W_{\text{new}} = W + \Delta W = W + A \cdot B,
\end{equation}
where \( A \in \mathbb{R}^{d \times r} \) and \( B \in \mathbb{R}^{r \times d} \) are low-rank matrices and \( r \ll d \). This decomposition retains expressiveness while ensuring training efficiency \cite{hu2022lora}.

In the second stage, only the intermediate semantic encoding $\mathbf{e}_{k,n}$ is transmitted to the aggregation center, where the TRAM module is trained. The aggregated loss is formulated as:
\begin{equation}\label{eq:TDL8}
    \mathcal{L}_\text{agg} = l_1\mathcal{L}_\text{d} + l_2\mathcal{L}_\text{a} + l_3\mathcal{L}_\text{p} + l_4\mathcal{L}_\text{v} + l_5\mathcal{L}_\text{i}.
\end{equation}

During the training process, each device receives a captured image $\mathbf{V}_k$ and an echo signal $\mathbf{R}_{k,n}$. Additionally, dynamic channel characteristics are incorporated, based on the SNR and the spatial distance between transmitters and receivers. These parameters, alongside contextual channel features, are used to condition the model predictions.
Assuming that the complete training dataset be denoted as $\mathcal{D} = \{\mathcal{D}_\text{agg}, \mathcal{D}_1, \dots, \mathcal{D}_K\}$, where $\mathcal{D}_\text{agg}$ is the dataset of the aggregation center, the overall training procedure for the LLM-DiSAC framework is summarized in \textbf{Algorithm~\ref{alg:TDL}}.
\begin{algorithm}
\caption{Training process based on two-stage distributed learning}
\label{alg:TDL}
\begin{algorithmic}[1]
    \REQUIRE $\mathcal{D}$.
    \ENSURE $\bm{\alpha}, \bm{\beta}, \bm{\gamma}, \bm{\delta}, \bm{\eta}, \bm{\zeta}, \bm{\epsilon}$.
    
    \vspace{0.5em}
    \noindent\textbf{First stage:}
    \FOR{each sensing device}
        \FOR{each batch $(\mathbf{V}_k, \mathbf{R}_{k,n})$ from $\mathcal{D}_k$ }
            \STATE{Generate communication parameters $C_{k,n}$ using dynamic SNR and distance.}
            \STATE{Predict the local sensing results $\hat{d}_{k,n}, \hat{v}_{k,n}, \hat{\phi}_{k,n}, \hat{\theta}_{k,n}, \hat{\mathbf{I}}_{k,n}$ according to \textbf{Algorithms \ref{alg:RVFN}-\ref{alg:LSTN}} and Eqs. (\ref{eq:TRAM5})-(\ref{eq:TRAM6}).}
            \STATE{Compute local training loss using Eqs. (\ref{eq:TDL1})-(\ref{eq:TDL6}).}
            \STATE{Backpropagate and update model parameters $\bm{\alpha}, \bm{\beta}, \bm{\gamma}, \bm{\delta}, \bm{\epsilon}$ with the optimizer.}
        \ENDFOR
    \ENDFOR

    \vspace{0.5em}
    \noindent\textbf{Second stage:}
     \FOR{each batch from $\mathcal{D}_\text{agg}$ }
    \STATE{Predict the sensing results $\mathbf{o}_{\text{agg},n}$ using \textbf{Algorithms \ref{alg:TRAM}}.}
    \STATE{Compute training loss using Eq. (\ref{eq:TDL8}).}
    \STATE{Backpropagate and update model parameters $\bm{\eta}, \bm{\zeta}$ with the optimizer.}
      \ENDFOR
\end{algorithmic}
\end{algorithm}

\section{Experimental Setup and Numerical Results}
This section presents the simulation dataset, parameter configurations, and evaluation results. The simulations are conducted on a server equipped with an Intel Xeon CPU (2.3 GHz, 256 GB RAM) and two NVIDIA RTX 4090 GPUs (24 GB SGRAM each), leveraging the PyTorch framework to implement the proposed schemes. 

\subsection{Experimental Settings}
\subsubsection{Dataset Setup}  
\begin{figure*}[htbp]
	\centering
	\includegraphics[width=18cm]{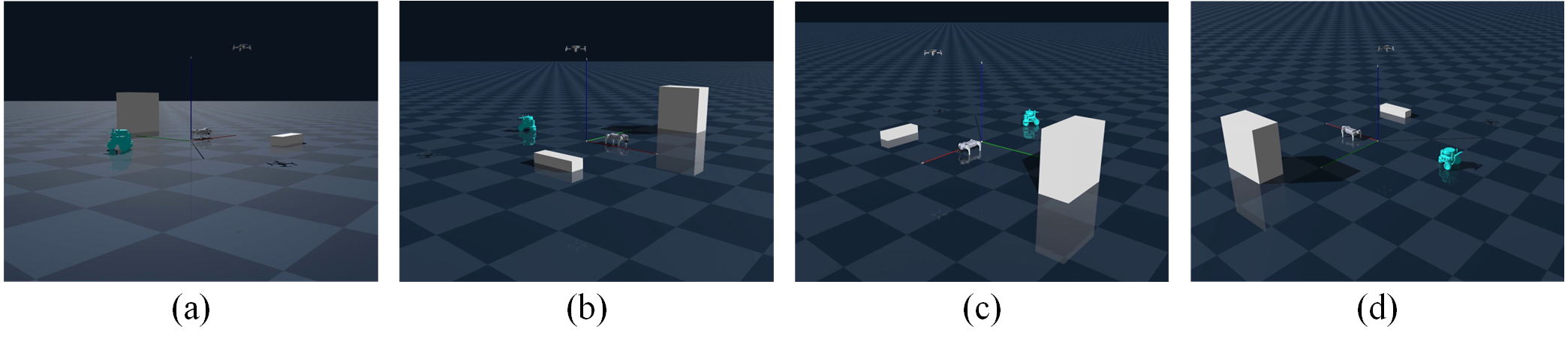}
	\caption{An example of the visual observations on: (a) position 1, (b) position 2, (c) position 3, and (d) position 4.}
	\label{fig:exp_vis}
\end{figure*}

To evaluate the efficacy of our proposed methods, we construct a synthetic multi-view RF-visual sensing dataset using Genesis \cite{Genesis}, an AI-powered physical engine. 
Specifically, we employ the Genesis simulation platform to construct a scenario, in which four sensing devices are deployed at different spatial locations and altitudes to observe a common region of interest. This region contains three distinct moving targets, namely, a car, a robotic dog, and a drone, in addition to several static obstacles. Each moving target performs uniform linear motion in random directions at a given velocity $v_{n}$, simulating dynamically diverse motion patterns. All four sensing devices simultaneously capture the scene's video footage and record the moving targets' real-time trajectories. For each device, a total of 37 video sequences are collected. Each sequence has a duration of 5 seconds and is recorded at a frame rate of 120 frames per second.
Fig. \ref{fig:exp_vis} presents an example of the visual observations of the scene captured from different positions. Next, we describe the construction of the synthetic multi-view RF-visual sensing dataset based on the collected data.

Firstly, each video is transformed into frames, resulting in approximately 22,200 ($37 \times 5 \times 120$) RGB images (i.e., $\mathbf{V}_k$).  
For each extracted frame, we calculate the distance of each ST to the sensing device as follows:
\begin{equation}  
    d_{k,n}=\sqrt{(x^\text{SD}_k-x_{n}^\text{center})^2+(y^\text{SD}_k-y_{n}^\text{center})^2},  
\end{equation}  
where $(x^\text{SD}_k, y^\text{SD}_k)$ represents the coordinates of the $k$th IoTD, and $(x_{n}^\text{center}, y_{n}^\text{center})$ denotes the center coordinates of the $n$th ST's bounding box. 

Secondly, we estimate the azimuth angle of each ST relative to the sensing device as follows:  
\begin{equation}  
    \phi_{k,n}=\operatorname{arctan2}(y_k^\text{SD}-y_{n}^\text{center}, x_{n}^\text{center}-x_k^\text{SD}),  
\end{equation}
where $\operatorname{arctan2}(\cdot)$ computes the arc tangent of the input coordinates, returning a radian value within the range $[-\pi, \pi]$. Similarly, the pitch angle can be calculated by:
\begin{equation}
\begin{split}
    \theta_{k,n} = \operatorname{arctan2}\Big(  
        z_k^\text{SD} - z_{n}^\text{center},\; \\ 
        \sqrt{(x_{n}^\text{center} - x_k^\text{SD})^2 + (y_{n}^\text{center} - y_k^\text{SD})^2}  
    \Big),
\end{split}
\end{equation}
where $z_k^\text{SD}$ and $ z_n^\text{center}$ represent the altitude of the $k$th IoTD and the $n$th ST, respectively.

Third, we estimate the radial velocity of the $n$th ST relative to the $k$th IoTD as follows:  
\begin{equation}
v_{k,n} =
\frac{
  \begin{array}{c}
    (x_{n}^\text{vel},\; y_{n}^\text{vel},\; z_{n}^\text{vel}) \cdot \\
    (x_{n}^\text{center} - x_k^\text{SD},\; y_{n}^\text{center} - y_k^\text{SD},\; z_{n}^\text{center} - z_k^\text{SD})
  \end{array}
}{
  \sqrt{
    (x_{n}^\text{center} - x_k^\text{SD})^2 +
    (y_{n}^\text{center} - y_k^\text{SD})^2 +
    (z_{n}^\text{center} - z_k^\text{SD})^2
  }
}
\end{equation}
where $(x_{n}^\text{vel}, y_{n}^\text{vel}, z_{n}^\text{vel})$ denotes the velocity vector of the $n$th ST, and the numerator represents the dot product between the velocity vector and the direction vector pointing from the sensing device to the ST. 

Next, when the $k$th IoTD is the aggregation center, we have $d_{\text{agg},n}=d_{k,n}$, $\phi_{\text{agg},n}=\phi_{k,n}$, $\theta_{\text{agg},n}=\theta_{k,n}$, $v_{\text{agg},n}=v_{k,n}$, and $\mathbf{I}_{\text{agg},n}=\mathbf{I}_{k,n}$.

Finally, we generate the RF signal $\mathbf{R}_{k,n}$ for each ST according to Eq. (\ref{eq:sig}). As a result, the training dataset $\mathcal{D}$ is obtained.

\subsubsection{Parameter Settings}  
In the system model, the AWGN channel is considered within the SC framework. The weighting factors for task-specific losses are set to \( l_1 = l_2 = l_3 = l_4 = 50 \) and \( l_5 = 1 \). The bandwidth is configured as \( B = 1 \, \text{kHz} \), with a transmission power of \( P = 1 \, \text{W} \), and the SNR varies from 0 dB to 25 dB. The radar operates at a carrier frequency of \( f_c = 10 \, \text{GHz} \), with a Pulse Repetition Interval (PRI) of \( T_r = 1 \times 10^{-6} \, \text{s} \). The sampling frequency is \( F_s = 60 \, \text{MHz} \), corresponding to a sampling interval of \( \Delta t = \frac{1}{F_s} = 1.67 \, \text{ns} \). The RCS for the three STs—car, robotic dog, and drone—are set to 100, 10, and 1, respectively. A SIMO radar model is employed, utilizing \(T=16 \) antennas to transmit LFM waveforms and receive echo signals.

During training, the SNR is randomly selected for each forward pass to enhance the robustness of the proposed LLM-DiSAC framework against channel noise. In the inference phase, the LLM-DiSAC is evaluated under fixed SNR conditions of 0, 10, 15, 20, and 25 dB.

\subsubsection{Benchmark Schemes}  
To evaluate the performance of the proposed LLM-DiSAC framework, we conducted experiments from three perspectives: \textit{(i) multimodal versus unimodal sensing, (ii) SC models based on traditional architectures versus those based on LLM, and (iii) single-device versus multi-device collaborative sensing.} The corresponding benchmark schemes for each experimental dimension are detailed as follows.

\textbf{Benchmarks for LLM-Based vs. Traditional Model-Based Sensing:}
\begin{itemize}  
    \item \textbf{LLM-DiSAC (with GPT-2)}: This variant integrates GPT-2 as the semantic decoder within the LSTN module.
    \item \textbf{LLM-DiSAC (with LSTM)}: This version employs an LSTM-based semantic decoder in the LSTN.
    \item \textbf{LLM-DiSAC (with GRU)}: This variant utilizes a GRU-based semantic decoder in the LSTN.
    \item \textbf{LLM-DiSAC}: Our proposed framework that incorporates the LLM-enhanced SC and sensing.
\end{itemize}

\textbf{Benchmarks for Multimodal vs. Unimodal and Multi-Device vs. Single-Device Sensing:
}
\begin{itemize}  
    \item \textbf{SM-SD (RF)}: A single-modal, single-device setting that employs only the RF feature extractor and the multi-head prediction network.
    \item \textbf{SM-SD (CV)}: A single-modal, single-device setting that utilizes only the CV feature extractor and the multi-head prediction network.
    \item \textbf{MM-SD}: A multi-modal, single-device setting that leverages the RVFN module and the multi-head prediction network.
    \item \textbf{SM-MD (RF)}: A single-modal, multi-device setting employing the RF feature extractor in conjunction with LSTN and TRAM.
    \item \textbf{SM-MD (CV)}: A single-modal, multi-device setting utilizing the CV feature extractor along with LSTN and TRAM.
    \item \textbf{LLM-DiSAC}: Our proposed scheme enables multi-modal, multi-device collaborative sensing.
    \item \textbf{LLM-DiSAC (w/o SC loss)}: A theoretical upper-bound baseline in which SC is assumed to be lossless between collaborative devices and the aggregation center.
\end{itemize}

\subsubsection{Evaluation Metrics}
We use the normalized mean squared error (NMSE) and the root mean squared error (RMSE) to evaluate the performance of the proposed method in distance, radial velocity, azimuth, and pitch angle prediction tasks. 
NMSE reflects the prediction accuracy by comparing the mean squared error with the signal power, and is defined as follows:
\begin{equation}
    \text{NMSE}(\mathbf{x}_i, \hat{\mathbf{x}}_i) = 10\log{\frac{\sum_{i=1}^I \| \hat{\mathbf{x}}_i - \mathbf{x}_i \|^2}{\sum_{i=1}^I \| \mathbf{x}_i \|^2}},
\end{equation}  
where \( \hat{\mathbf{x}}_i \in (0, 1) \) denotes the predicted value, \( \mathbf{x}_i \) is the ground truth, and \( I \) is the total number of samples. A smaller NMSE indicates better prediction performance.
RMSE quantifies the absolute average deviation between predictions and ground truth. These metrics are defined as follows:  
\begin{equation}  
    \text{RMSE}(\mathbf{x}_i,\hat{\mathbf{x}}_i) = \sqrt{\frac{1}{I} \sum_{i=1}^I \| \hat{\mathbf{x}}_i - \mathbf{x}_i \|^2}. 
\end{equation}  

We adopt classification accuracy to evaluate the LLM-DiSAC performance on identifying the class of each ST. Accuracy measures the proportion of correctly predicted samples among the total number of samples, and is defined as follows:  
\begin{equation}
    \text{Accuracy} = \frac{1}{I} \sum_{i=1}^I \mathbb{1}(\hat{y}_i = y_i),
\end{equation}  
where \( \hat{y}_i \) and \( y_i \) denote the predicted and ground-truth labels of the \( i \)th sample, respectively, \( I \) is the total number of samples, and \( \mathbb{1}(\cdot) \) is the indicator function that equals 1 if the input condition is true and 0 otherwise. A higher accuracy indicates better classification performance.

\subsection{Evaluation Results}
Since the proposed LLM-DiSAC framework is designed to enhance sensing accuracy through a distributed collaborative architecture and the modeling capacity of LLM, we primarily assess its performance by conducting comparative experiments, focusing on the impact of collaboration and the LLM component.

\subsubsection{Evaluation for LLM-Based vs. Traditional Model-Based Sensing}
\begin{figure*}[htbp]
	\centering
	\includegraphics[width=18cm]{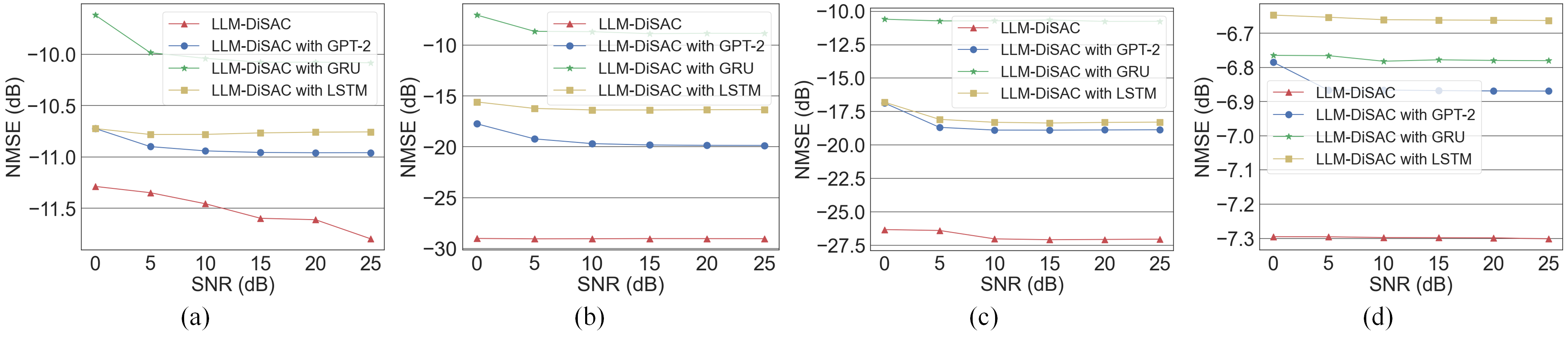}
	\caption{NMSE results under the AWGN channel under different motion parameters: (a) distance, (b) azimuth, (c) pitch, and (d) velocity.}
	\label{fig:exp_cp_nmse_awgn}
\end{figure*}
\begin{figure*}[htbp]
	\centering
	\includegraphics[width=18cm]{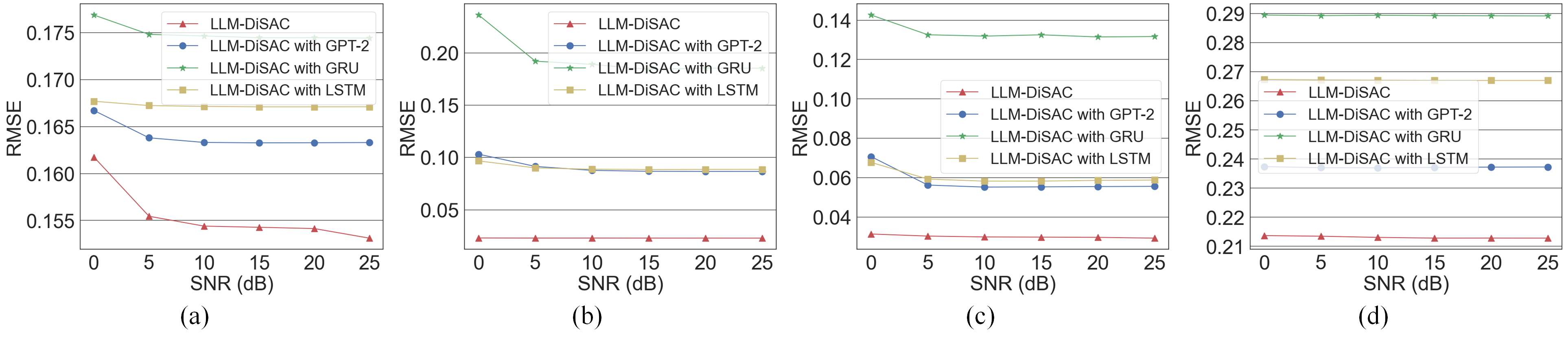}
	\caption{RMSE results under the AWGN channel under different motion parameters: (a) distance, (b) azimuth, (c) pitch, and (d) velocity.}
	\label{fig:exp_cp_rmse_awgn}
\end{figure*}
\begin{figure}[htbp]
	\centering
	\includegraphics[width=8cm]{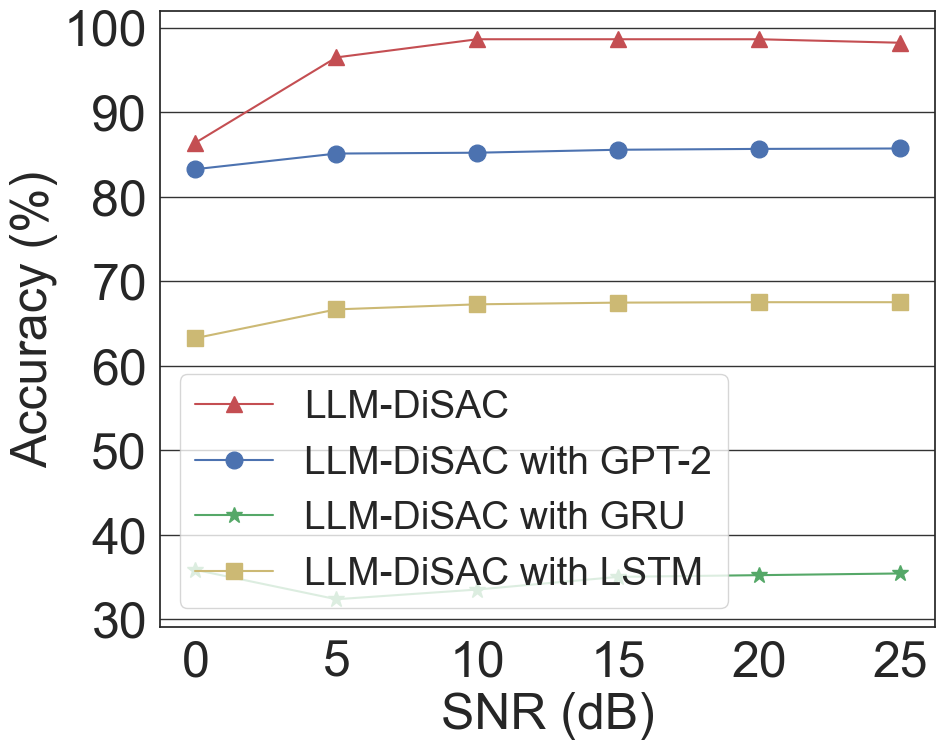}
	\caption{Classification accuracy on different schemes under different SNRs.}
	\label{fig:exp_cp_cls}
\end{figure}
We evaluate the performance of the proposed LLM-DiSAC framework and its baseline variants, LLM-DiSAC with GPT-2, GRU, and LSTM, under varying SNR levels, with the aggregation center fixed at position 1. Fig. \ref{fig:exp_cp_nmse_awgn} presents the NMSE results across four key motion parameters: distance, azimuth, pitch, and velocity, respectively. All baseline variants achieve consistently lower NMSE than the original LLM-DiSAC, particularly at higher SNRs. Among them, LLM-DiSAC with GPT-2 frequently delivers superior performance, especially in estimating azimuth, pitch, and velocity, underscoring their effectiveness in capturing complex motion dynamics in noise-dominated scenarios. These trends are further supported by Fig. \ref{fig:exp_cp_rmse_awgn}, which presents the RMSE results and corroborates the findings from the NMSE analysis.

Fig. \ref{fig:exp_cp_cls} shows the classification accuracy of the LLM-DiSAC and its baselines. The default LLM-DiSAC, which integrates the LLaMA-3 model, consistently achieves the highest classification accuracy across all SNR levels. All models benefit from increasing SNR, with the GPT-2-enhanced LLM-DiSAC ranking second in performance. These results highlight the critical importance of model capacity in semantic information recovery, demonstrating that LLMs like LLaMA-3 and GPT-2 significantly outperform smaller architectures across diverse channel conditions.

Additionally, in our experiments, transmitting a set of visual and radio frequency data using the conventional method requires 1,370,112 bytes, whereas the proposed SC scheme reduces this to only 100,352 bytes, achieving an approximate 92.67\% reduction in transmission size.

In summary, integrating LLaMA-3 into the LLM-DiSAC framework can fully utilize the channel parameters and recover the latent information from the received semantic encoding, effectively bridging textual and non-textual modalities, improving transmission efficiency and downstream feature aggregation performance in SC systems.

\subsubsection{Evaluation for Multimodal vs. Unimodal and Multi-Device vs. Single-Device Sensing}

\renewcommand{\arraystretch}{1.3} 
\begin{table*}[htbp]
\caption{Comparison results of multimodal vs. unimodal and multi-device vs. single-device sensing}
\label{tab:exp}
\centering
\begin{tabular}{lccccccccc}
\toprule
\textbf{Method} & \begin{tabular}[c]{@{}c@{}}NMSE of \\ distance (dB)\end{tabular} & \begin{tabular}[c]{@{}c@{}}RMSE of \\ distance\end{tabular} & \begin{tabular}[c]{@{}c@{}}NMSE of \\ azimuth (dB)\end{tabular} & \begin{tabular}[c]{@{}c@{}}RMSE of \\ azimuth\end{tabular} & \begin{tabular}[c]{@{}c@{}}NMSE of \\ pitch (dB)\end{tabular} & \begin{tabular}[c]{@{}c@{}}RMSE of \\ pitch\end{tabular} & \begin{tabular}[c]{@{}c@{}}NMSE of \\ velocity (dB)\end{tabular} & \begin{tabular}[c]{@{}c@{}}RMSE of \\ velocity\end{tabular} & \begin{tabular}[c]{@{}c@{}}Classification\\ accuracy (\%)\end{tabular} \\
\midrule
SM-SD (RF)                  & -11.442 & 0.166 & -24.300 & 0.035 & -21.344 & 0.042 & -6.659 & 0.284 & 91.60 \\
SM-SD (CV)                  & -9.237  & 0.168 & -26.194 & 0.026 & -9.925  & 0.146 & -6.518 & 0.287 & 33.75 \\
MM-SD                       & -11.298 & 0.166 & -26.788 & 0.027 & -24.305 & 0.029 & -6.736 & 0.237 & 92.07 \\
SM-MD (RF)                  & -11.549 & 0.157 & -28.694 & 0.023 & -26.751 & 0.029 & -7.102 & 0.214 & 97.02 \\
SM-MD (CV)                  & -9.313  & 0.167 & -28.642 & 0.023 & -12.774 & 0.141 & -7.121 & 0.213 & 36.67 \\
LLM-DiSAC                   & -11.699 & 0.154 & -29.060 & 0.023 & -27.046 & 0.029 & -7.302 & 0.212 & 98.21 \\
\begin{tabular}[c]{@{}l@{}}LLM-DiSAC\\ (w/o SC loss)\end{tabular} & -11.970 & 0.153 & -29.758 & 0.022 & -27.882 & 0.029 & -7.339 & 0.204 & 99.80 \\
\bottomrule
\end{tabular}
\end{table*}

Table \ref{tab:exp} presents the performance comparison across different configurations involving unimodal vs. multimodal and single-device vs. multi-device sensing.
Several key observations can be made. First, multimodal sensing consistently outperforms unimodal settings in both regression and classification tasks. For example, MM-SD achieves significant improvements over both SM-SD (RF) and SM-SD (CV) in terms of NMSE and RMSE across all sensing dimensions, as well as a notable increase in classification accuracy (92.07\% vs. 91.60\% and 33.75\%, respectively). 
Quantitatively, multimodal sensing improves classification accuracy by up to 172.7\% and reduces RMSE by an average of 15.8\% and NMSE by an average of 5.7\% across distance, pitch, and velocity dimensions.
This highlights the complementary nature of RF and CV modalities, where fusing information from different sources enhances the model’s robustness and expressiveness.

Second, incorporating multi-device collaboration further boosts performance. Comparing SM-SD (RF) to SM-MD (RF), and SM-SD (CV) to SM-MD (CV), we observe that the use of multiple devices, enabled by modules like LSTN and TRAM, leads to lower NMSE/RMSE and higher classification accuracy. 
Specifically, multi-device collaboration leads to a classification accuracy gain of 5.9\% for RF-based methods (from 91.60\% to 97.02\%) and 8.7\% for CV-based methods (from 33.75\% to 36.67\%), while also reducing RMSE by an average of  18.73\% and NMSE by an average of 11.7\% across distance, pitch, and velocity dimensions.
This indicates that spatially distributed sensing provides richer and more diverse observations, thereby improving the fidelity of perception tasks.

Finally, our proposed method, LLM-DiSAC, which combines multimodal inputs with multi-device collaboration, achieves the best overall performance. Compared to the weakest baseline (SM-SD (CV)), LLM-DiSAC achieves a 191.0\% relative improvement in classification accuracy, reduces RMSE by an average of 31.5\%, and improves NMSE by an average of 55.6\%, demonstrating comprehensive advantages in both regression and classification tasks.
This demonstrates that both multimodal fusion and multi-device collaboration are critical for enabling high-precision and reliable collaborative sensing. Additionally, LLM-DiSAC (w/o SC loss) provides an upper bound (e.g., classification accuracy of 99.80\%), validating our proposed gap between our method and the theoretical upper.

\begin{figure*}[htbp]
	\centering
	\includegraphics[width=18cm]{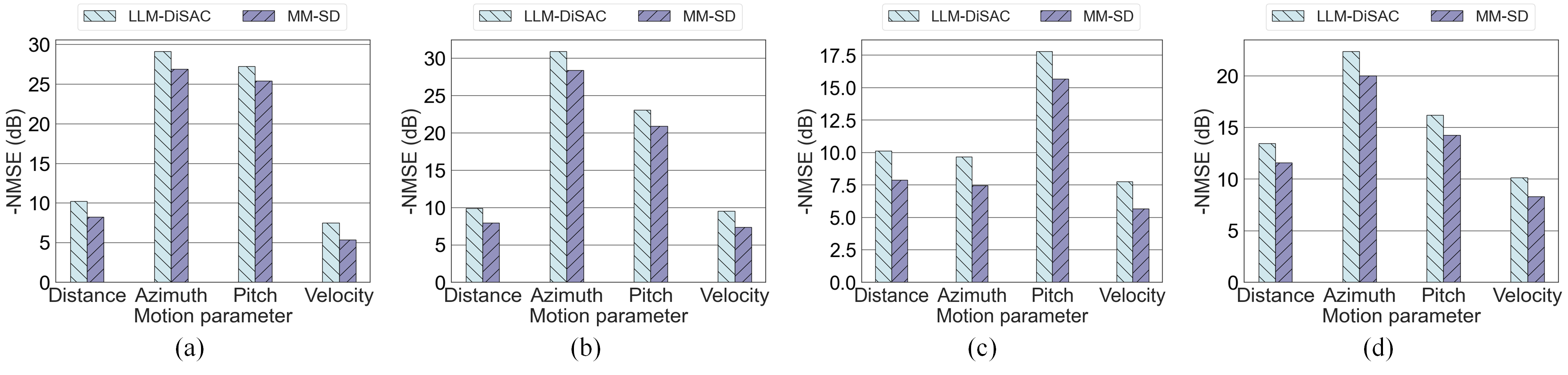}
	\caption{NMSE results under the AWGN channel when the aggregation center is on (a) position 1, (b) position 2, (c) position 3, and (d) position 4, respectively.}
	\label{fig:exp_nmse_awgn}
\end{figure*}
\begin{figure*}[htbp]
	\centering
	\includegraphics[width=18cm]{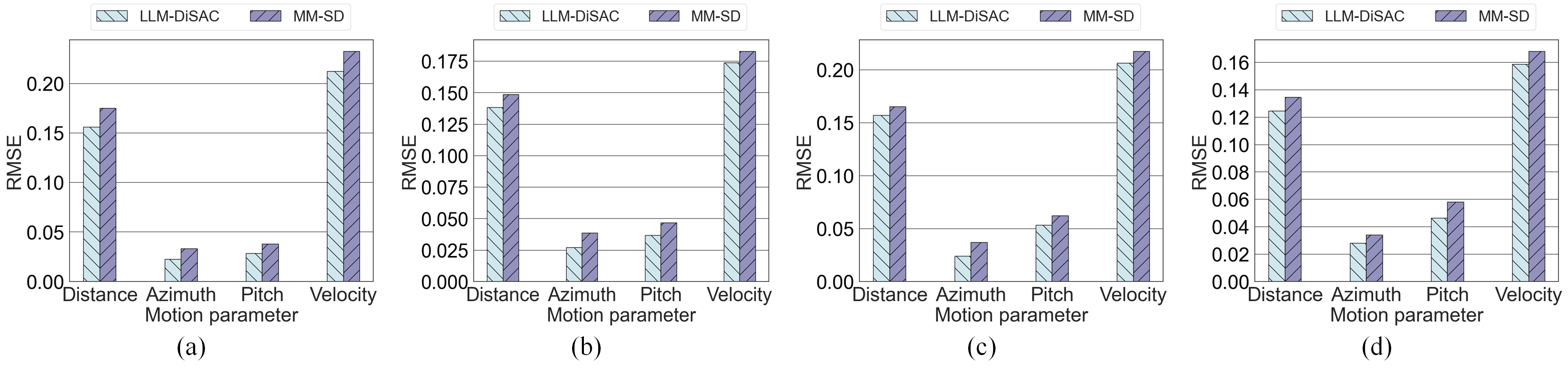}
	\caption{RMSE results under the AWGN channel when the aggregation center is on (a) position 1, (b) position 2, (c) position 3, and (d) position 4, respectively.}
	\label{fig:exp_rmse_awgn}
\end{figure*}
\begin{figure}[htbp]
	\centering
	\includegraphics[width=8cm]{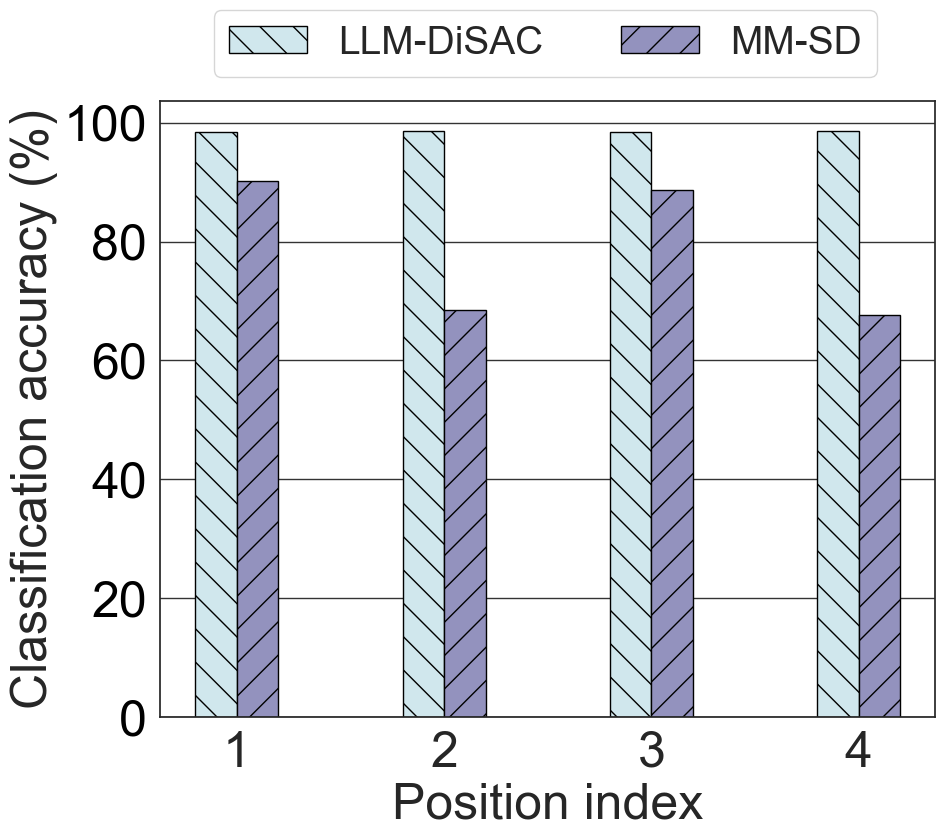}
	\caption{Classification accuracy when the aggregation center is in different positions.}
	\label{fig:exp_cls}
\end{figure}

We also evaluate the performance of LLM-DiSAC and MM-SD under different aggregation center positions. Fig. \ref{fig:exp_nmse_awgn} presents the NMSE performance. In both scenarios, results are reported for four distinct aggregation center positions. It is evident that LLM-DiSAC consistently outperforms MM-SD across all motion parameters, distance, azimuth, pitch, and velocity. This demonstrates the effectiveness of the TRAM module in enhancing motion parameter estimation. The RMSE results in Fig. \ref{fig:exp_rmse_awgn} further corroborate these findings, exhibiting similar trends as those observed in the NMSE evaluations. Fig. \ref{fig:exp_cls} shows the classification accuracy of LLM-DiSAC and MM-SD across varying positions of the aggregation center, where LLM-DiSAC maintains consistently high classification accuracy, with only minor fluctuations across positions. In contrast, MM-SD exhibits more noticeable positional sensitivity, suggesting that fixed-angle sensing configurations can introduce variance in recognition performance.

Overall, the advantage of LLM-DiSAC is particularly evident in angular motion parameters such as azimuth and pitch, as well as target identification. Furthermore, variations in estimation accuracy with different aggregation center positions suggest that device placement plays a non-trivial role in sensing effectiveness. These results validate the contribution of the TRAM module, which, through its cross-aggregation attention mechanism, enables the aggregation center to adaptively fuse complementary sensing features from multiple devices, thereby overcoming the limitations of a single device with a fixed viewing angle and significantly improving sensing accuracy.

\section{Conclusions and Future Works}
In this paper, we propose LLM-DiSAC, a novel LLM-driven framework for distributed multimodal sensing and SC, which addresses key limitations in traditional systems such as limited modality, restricted sensing coverage, and inefficient feature aggregation. By integrating RF and visual data via the RVFN module with cross-attention, the framework effectively overcomes the constraints of single-modal sensing. TRAM adaptively fuses features from multiple devices to extend the perception range beyond fixed fields of view, while LSTN enables efficient SC by leveraging physical priors and enhancing decoding robustness. A two-stage distributed learning strategy is further introduced to ensure data privacy. Experimental results on a synthetic multi-view RF-visual dataset demonstrate that LLM-DiSAC achieves a 191.0\% relative improvement in classification accuracy, reduces RMSE by an average of 31.5\%, and lowers NMSE by 55.6\%, compared to unimodal single-device baselines. Moreover, the data transmission cost is reduced by 92.6\% relative to conventional approaches. These results underscore the comprehensive advantages of multimodal fusion, multi-device collaboration, and SC.

In the future, we will focus on bridging the gap between simulation and real-world deployment. As the current study is based on synthetic data generated by a physics-driven simulator, our next step is to build a practical experimental platform to collect real-world RF-visual data under diverse environmental and channel conditions. This will allow for comprehensive validation of the LLM-DiSAC framework’s robustness and adaptability in practical scenarios, thereby further demonstrating its potential for real-world applications in edge intelligence.

\bibliographystyle{IEEEtran}
\bibliography{bare_jrnl}
\newpage
\end{document}